\documentclass[twocolumns]{aa} % for a paper on 1 column  
\usepackage{graphicx}
\usepackage{float}
\usepackage{subfig}
\usepackage{txfonts}
\usepackage{comment}
\usepackage{url}
\usepackage[switch]{lineno}
\usepackage{datetime}
\usepackage[usenames,table,dvipsnames]{xcolor}
\usepackage{setspace}
\definecolor{blue}{RGB}{66, 153, 233}
\definecolor{red}{RGB}{255, 0, 0}
\usepackage[normalem]{ulem}

\newcount\Comments  % 0 suppresses notes to selves in text
\newcommand{\kibitz}[2]{\ifnum\Comments=1\textcolor{#1}{#2}\fi}

\begin{document}

   \title{Novel method for component separation of extended sources in X-ray astronomy}

   \author{A. Picquenot\inst{1}
          \and
          F. Acero\inst{1}
          \and
          J. Bobin\inst{1}
          \and 
          P. Maggi\inst{2}
          \and 
          J. Ballet\inst{1}
                \and 
          G.W. Pratt\inst{1}}
 
          %\fnmsep\thanks{Just to show the usage
          %of the elements in the author field}

   \institute{AIM, CEA, CNRS, Universit\'e Paris-Saclay, Universit\'e Paris Diderot, Sorbonne Paris Cit\'e, F-91191 Gif-sur-Yvette, France
              \email{adrien.picquenot@cea.fr, fabio.acero@cea.fr}
              \and
              Observatoire Astronomique de Strasbourg, Universit\'e de Strasbourg, CNRS, 11 rue de l'Universit\'e, F-67000 Strasbourg, France
             }
             
%   \institute{Strasbourég             }
             
   \date{\today}

 \abstract{     In high-energy astronomy, spectro-imaging instruments such as X-ray detectors allow investigation of the spatial and spectral properties of extended sources including galaxy clusters, galaxies, diffuse interstellar medium, supernova remnants, and pulsar wind nebulae. In these sources, each physical component possesses a different spatial and spectral signature, but the components are entangled. Extracting the intrinsic spatial and spectral information of the individual components from this data is a challenging task. Current analysis methods do not fully exploit the 2D-1D ($x,y,E$) nature of the data, as spatial information  is considered separately from spectral information. Here we investigate the application of a blind source separation (BSS) algorithm that jointly exploits the spectral and spatial signatures of each component in order to disentangle them. We explore the capabilities of a new BSS method
(the general morphological component analysis; GMCA), initially developed to extract an image of the Cosmic Microwave Background from {\it Planck} data, in an X-ray context. The performance of the GMCA on X-ray data is tested using Monte-Carlo simulations of supernova remnant toy models designed to represent typical science cases. We find that the GMCA is able to separate highly entangled components in X-ray data even in high-contrast scenarios, and can extract the spectrum and map of each physical component with high accuracy. A modification of the algorithm is proposed in order to improve the spectral fidelity in the case of strongly overlapping spatial components, and we investigate a resampling method to derive realistic uncertainties associated to the results of the algorithm. Applying the modified algorithm to the deep \textit{Chandra} observations of Cassiopeia A, we are able to produce detailed maps of the synchrotron emission at low energies (0.6-2.2 keV), and of the red- and blueshifted distributions of a number of elements including  Si and Fe K.}


   \maketitle
%
%-------------------------------------------------------------------
\section{Introduction}
\label{intro}

    %Several astronomical objects can be described as extended sources emitting in X-rays, among which we find galaxy clusters, pulsar wind nebulae or supernova remnants (SNRs). 
    Beginning in the 1970s, it was realised that the X-ray sky is full of extended sources, among which we find emission from the Milky Way itself, other Galactic sources such as pulsar wind nebulae or supernova remnants (SNRs), and extragalactic sources such as galaxies and clusters of galaxies. The typical  emission components one can see in X-rays from these types of objects are thermal emission or accelerated particles radiating through the synchrotron process. In each case, their spectral signature is distinctive and recognizable. For example, in SNRs the shock wave propagating rapidly through the interstellar medium heats it up to approximately $10^7$ K, resulting in thermal emission peaking in the X-ray domain.
    
        Spectro-imaging instruments such as those aboard the current generation of X-ray satellites {\it XMM-Newton} and {\it Chandra} provide data comprising spatial and spectral information: the detectors record the position $(x,y)$ and energy $E$ event by event, thereby providing a data cube with two spatial dimensions and one spectral dimension.
    %\pierre{Also a spatial dimension, although it might not be relevant here if we do not explore time variability.} \pierre{I meant of course a \textbf{time} dimension} 
    
    An ability to disentangle the different physical components in this 2D-1D data cube would allow us to learn more about their respective spatial and spectral distributions. However, the different components are frequently superimposed along the line of sight, or are even physically nested, making such separation difficult.%are overlapping along the line of sight or even physically nested, making it a hard task to separate them properly.
    
        In this paper we introduce a new method to disentangle spectral components from X-ray data of extended sources. Separating a set of components mixed in a set of observations is known in the field of signal processing as a blind source separation (BSS) problem. Our method is based on an algorithm that uses the ability of wavelets to provide a sparse representation for astrophysical images to find a solution to BSS problems. In this context, we consider our 2D-1D data cube as the product between an image and a spectrum. This algorithm, the generalized morphological components analysis (GMCA), was first developed by \citet{bobin15}, and has recently been applied to {\it Planck} survey data to separate the image of the Cosmic Microwave Background (CMB) from the foregrounds \citep{bobin16}. The application of the GMCA method to X-ray data is nontrivial. While in {\it Planck} the data are obtained in nine fixed frequencies, the X-ray photons can be binned into an arbitrarily large number of energy bins; the X-ray photon count is drastically lower at high energies, and has higher dynamic range. In addition, the X-ray data have Poisson noise whereas the GMCA method assumes an additive Gaussian noise.
%    However, X-rays and radio data present dissimilarities: there are more than a hundred energy channels in X-rays, compared to nine in Planck, the statistics are drastically lower at high energies with high dynamic range and the noise follows a Poisson rule instead of an additive Gaussian noise. "
    
        Here we adapt the GMCA algorithm to the study of extended sources in X-rays, and test its implementation by applying the method to SNR data. We first test the method on toy models reproducing X-ray data of SNRs containing up to three components (see Sections \ref{toy} and \ref{newtoys}). Although the noise is Poissonian in our simulated data set, we obtain accurate spectral shapes and cleaner images than with any of the typical X-ray analysis techniques. However, we find that a strong spatial correlation between the components leads to a leakage from the main components to the weaker ones, which may be partially linked to the nature of the noise, and we implement a refinement step in the algorithm to minimize this effect (see Section \ref{inpainting}). Although a version of the GMCA handling Poisson statistics is currently being developed, our results show that the existing version can be used to disentangle extended sources in X-rays. Applying our method to real data from the Cassiopeia A SNR yields sharp images of the synchrotron at low and high energy, and images of the distributions of a number of elements including Si and Fe K (see Section \ref{sect:real data}). In both cases, one of the images presents the blueshifted part of the structure, and the other one the redshifted part.

%--------------------------------------------------------------------
\section{Motivations and current methods}
\label{motivation}

The telescopes \textit{XMM-Newton} and \textit{Chandra} have provided a major step forward in effective area and angular resolution, and have led to nearly 20 years of observations resulting in deep (Mega-seconds) archival public datasets. As an example, the deep \textit{Chandra} $\sim$2 Ms observation of SNR Cassiopeia A resulted in about a billion X-ray photons. Despite this breakthrough improvement in data quality, the analysis techniques used to extract the wealth of information contained in such datasets have stalled.

  The main analysis challenge lies in the fact that at each position, the different spectral components (e.g., synchrotron and thermal emission from the shocked medium and ejecta) are projected along the line of sight, and that the observed signal is a combination of these components.
 
        In the study of SNRs, a typical scientific case is to study the spatial distribution of a spectral feature (e.g., heavy element maps to probe the morphology and asymmetries of the ejecta). The common methods are to generate maps integrated around the centroid energy of a line and to subtract the underlying continuum estimated from adjacent energy bands (the continuum interpolation method). However, if the faint emission lines are dominated by the continuum or if the adjacent energies also have emission lines, those methods perform poorly. An alternative method to study the spatial variations of the spectral properties is to divide the image into subregions and carry out a spectral analysis in each subregion. One frequently used method is to define regions of equal photon statistics with for example the Voronoï tiling method \citep[see][for an adaptation to X-rays]{diehl06b}. Each cell is then fitted with a physical model independently from its neighbors and maps representing the best-fit parameters are produced. This method is time consuming and does not take into account the underlying relationship between the spatial and spectral components. In addition, the best-fit parameter map may suffer from statistical fluctuations from cell to cell as for practical reasons only one grid is defined using a reference image in a large energy range that might not represent the flux of individual spectral components.

To summarize, one of the root issues of the methods described above is that each region (pixels or cell) is treated independently.  The disentangling process only relies on the spectral signature of the components in each region considered, whereas in reality the physical components also have different spatial signatures. Exploiting both the spectral and spatial signatures of the components and treating pixels not individually but as a whole yields more discriminative power to disentangle the different physical components. We note that other methods such as the principal component analysis (PCA) have already been applied to SNRs in the past to retrieve entangled components  \citep[see][]{Warren2005}. However, the PCA works in such a way that it has to retrieve decorrelated components, which usually makes them not physically significant. We can also cite \cite{2015ApJ...808..137J}, who used Bayesian statistical methods to infer the number of sources and probabilistically separate photons among the sources. Yet, these methods work with event lists $(x,y,E)$, and do not retrieve images or spectra associated with the sources, as our method does.
%       In high energy astrophysics, separating components is a key step in understanding several objects or mechanisms, but it has usually been handled poorly. The usual techniques are spectral interpolation and integration on a given energy range. Even if the images obtained this way can be correct for the main components, they are still polluted by the weaker ones, and these cannot be disentangled properly.

%--------------------------------------------------------------------
\section{A blind source separation method: the GMCA}
\label{GMCA}
\subsection{Description of the method}

\begin{figure*}[ht!]
\centering
\includegraphics[width = 19cm]{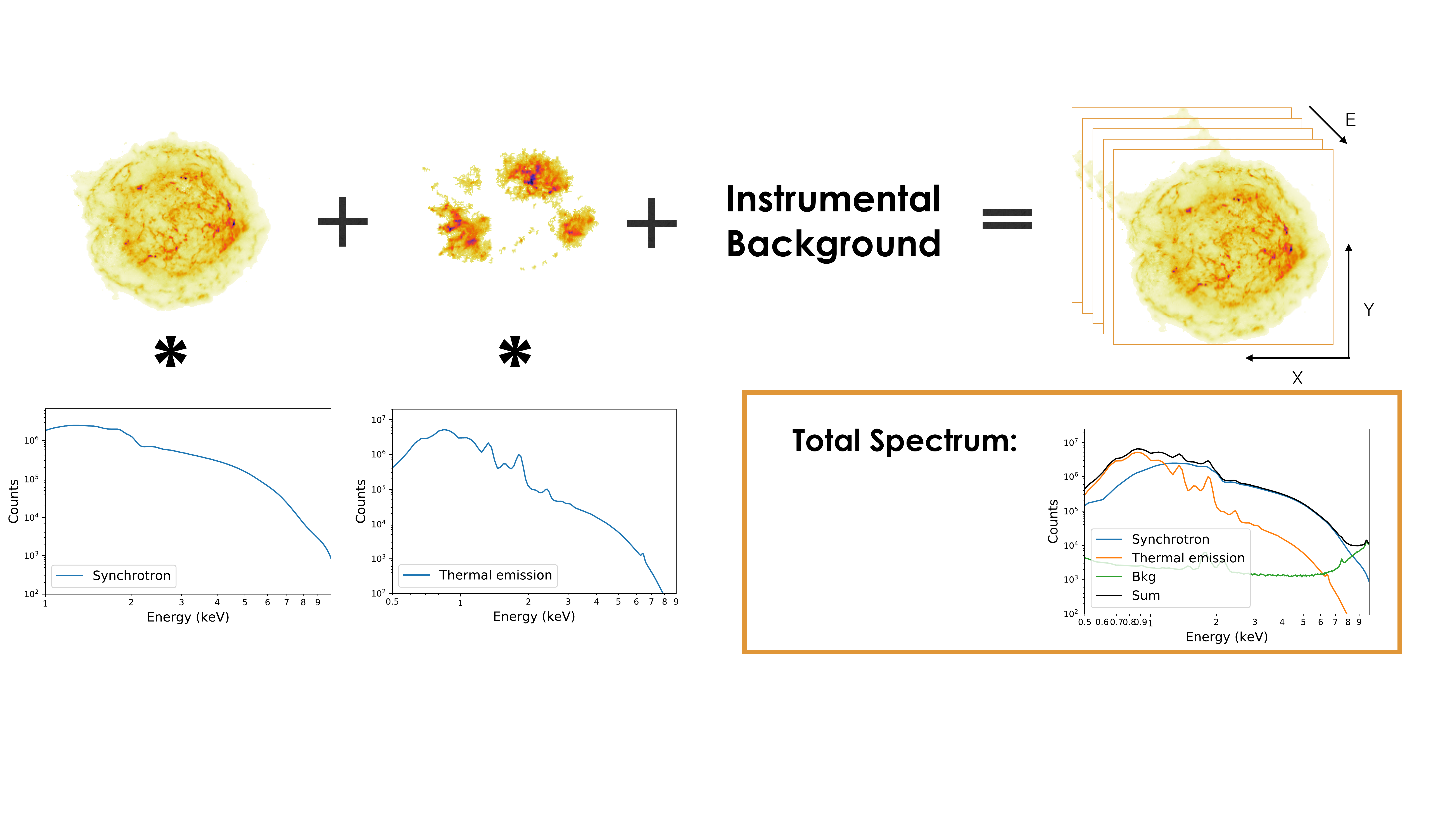}
%\vspace{-2.5cm}
\caption{A presentation of our toy model, consisting in the sum of images multiplied by theoretical spectra. The spatial distribution of the instrumental background is uniform.}
\label{fig:toy_model}
\end{figure*}

        Blind source separation methods aim to disentangle mixed sources in a data set without prior information. A classic way to do so is to look at the original data in a mathematical space where the sources will be sufficiently different from one another. The concept of sparsity helps to determine what kind of space could be suitable; a sparse signal is a signal in which most of the coefficients are zero. Thus, two sparse signals will be easier to disentangle as their signatures will not be correlated. For example, to separate periodic signals in a unidimensional data set, it is much easier to work in Fourier space, where such sources will be entirely determined by a few coefficients. 

         In this paper, we introduce a new method to disentangle physical components based on their spatial and spectral signature. This method is based on the GMCA algorithm, a blind source separation algorithm developed to disentangle the CMB from the galactic foregrounds in the data of the {\it Planck} satellite \citep{bobin15}. The input is a data cube $(E,x,y)$, where $E$ is the spectral dimension and $x$ and $y$ are spatial dimensions.

        The main concept of GMCA is to take into account the morphological particularities of each component to disentangle them. Apart from the $(E,x,y)$ data cube, the only input needed is the number $n$ of components to retrieve, which is user-defined. To optimize the disentangling process, the signal is projected in a space where it will have a sparse representation. Thus, two components that are sufficiently spatially different will have few coefficients in common, allowing us to separate them more easily. In the case of images, the equivalent of the Fourier space would be a correctly chosen wavelet transform, that would concentrate most of the image information into a few coefficients (for more about wavelets, and for an illustration of the interest of the wavelet space to disentangle components in a data cube $(E,x,y)$, see Appendix \ref{sect:wavelets}).
    
    %The sparsity we are looking for in our representation can be compared to that of a Fourier transformation in the case of an unidimensional periodic signal. Instead of the infinite number of non-zero coefficients necessary to describe it in the real space, a sinusoid can be entirely described by one coefficient in the Fourier space, and any periodic signal can be mostly described by a few coefficients in this same space. A sum of sinusoids will have a sparse representation in Fourrier space and a BSS algorithm will easily separate the individual components. In the case of images, wavelets can be adapted representations as few coefficients can concentrate most of the information.

\subsection{Mathematical formalism}

    Here we use the undecimated\footnote{An undecimated transform produces images of the same size for each scale.} Starlet transform \citep[see ][and Appendix \ref{sect:wavelets}]{starck2007} which is well suited for astronomical purposes. Each wavelet scale contains information about structures of a specific size, which allows us to isolate the morphological features of each component more easily. In order to minimize cross-correlations between components, the two largest wavelet scales are not used, because in these scales morphological features are harder to differentiate.
    
For a data cube of dimension $(E,x,y)$, we apply a wavelet transform with $J$ scales on the images of each energy slice of the cube resulting in an array $X$ of dimension $(E,x,y,J-2)$, the two largest wavelets scales being rejected. We note that the wavelet transform is applied only on images, and that there is no constraint on the sparsity of the spectra. The aim of the GMCA is to solve the following problem: 
%\jerome{Pas correct, il faut decoupler le modele que l'on fit aux donnees : $X = AS + N$ et le probleme resolu par GMCA. Ce dernier correspond a l'estimateur de A et S que l'on construit par minimzation des MC sous contrainte de parcimonie.}
\begin{equation}
\centering
X=AS+N=\overset{n}{\underset{i=1}{\sum}}A_i S_i+N
%Tips (Pierre): For sums and integral the proper notation is \sum\limits_{i=1}^{n}
,\end{equation}    

where $n$ is the predefined number of components, the $A_i$ are vectors of size $E$, in our case related to the spectral information (the spectra of our mixed components), the $S_i$ are the sources represented in wavelets, of dimension $(x,y,J)$ and related to the spatial information (the images in wavelets of our mixed components), and $N$ is a Gaussian noise. The product here for a given $i$ is the multiplication of every coefficient of $A_i$ by every coefficient of $S_i$. The components to retrieve are assumed to be modeled as the product of an image ($S_i$ in the wavelet space) and a spectrum ($A_i$). Thus, the retrieved components are approximations of the actual components with the same spectrum on each point of the image. This problem being an ill-posed inverse problem, as both $A$ and $S$ are unknown, one needs a constraint to solve it. 
%\gwp{La phrase suivante est une phrase clé, il faut qu'elle soit claire; elle ne l'est pas encore}The GMCA relies on the assumption that represented in a good basis (for example an undecimated Starlet basis), each source  can be sparsely represented, so that they can be separated. 
The GMCA relies on the assumption that once the image has been translated into wavelet space, each constituent can be sparsely represented, thus making the component separation easier.

        The GMCA solves the inverse problem by imposing a sparsity constraint: it maximizes the sparsity of the images of each source in the wavelet domain. The problem being actually solved by the GMCA is thus the following optimization problem:

\begin{equation}
\centering
\underset{A,S}{min} \overset{n}{\underset{i=1}{\sum}} \lambda_i \| S_i \|_p + \| X-AS \|_F^2
,\end{equation}

%\jerome{Il faut detailler l'equation qui est resolue}

        where $\lambda_i$ are regularization coefficients equivalent to thresholds that aim at rejecting noise samples, and are  essential to provide robustness with respect to noise. They are chosen thanks to an estimation of the noise level in the sources based on the median absolute deviation (MAD) method, and progressively decrease towards the final noise-related level.  $\| . \|_F$ is the Frobenius norm defined by $\| Y \|_F^2=$Trace$(YY^T)$ and $\| . \|_p$ is a $l_p$ norm, with $p=0$ or $p=1$. The $l_1$ norm is defined by $\| Y \|_1=\sum_{i,j} |Y_{i,j}|$ and $\| Y \|_0$ counts the number of nonzero entries in $Y$. The $l_0$ and $l_1$ norms are customarily used to measure the sparsity of signals. The first term of this equation is a sparsity constraint term and the second is a data-fidelity term. 
    
    More precisely, the GMCA is an iterative algorithm repeating the following two steps:
    
\begin{itemize}
\item Step 1: Estimation of $S$ for fixed $A$, by simultaneously minimizing $\| X-AS \|_F$ and the term enforcing sparsity in the Wavelet domain;
\item Step 2: Estimation of $A$ for fixed $S$ by minimizing $\| X-AS \|_F$.
\end{itemize}

\subsection{Application of the method}

    When the GMCA was applied to {\it Planck} data, the CMB spectrum was fixed to its theoretical shape. Giving a known spectrum as additional information fixes a column in $A$, making the algorithm work in what is termed a {\it semi-blind} mode. However, if the theoretical spectrum is not previously known, the algorithm can also work in a completely {\it blind} mode. With our toy model example (described in the following section), we test both of these modes.

The only input needed is the number $n$ of components to retrieve. Any prior knowledge of the data can help to choose $n$ wisely, that is, as the expected number of components visible in the energy band on which the GMCA is applied. In addition, this algorithm runs quickly (a few minutes to extract sources from a 200*200*300 single-core personal computer), so we highly recommend trying different values of $n$ and checking if the outputs have a physical relevance: as we see in Section \ref{imagefidelity}, the GMCA does not produce images of spurious structures.

We see in Appendix \ref{sect:numbcomp} that the Akaike information criterion (AIC) can be used as a figure of merit to confirm the relevance of a chosen $n$. However, this criterion must be used with caution since it is rigorously valid when computed at the maximum likelihood. This does not perfectly hold true in this case since: i) the underlying cost function that GMCA minimizes contains an additional sparsity regularization, and ii) the resulting problem is not convex and only a local minimizer is guaranteed to be reached.

%In theory, the number of components $n$ could be estimated along with the mixing matrix and the sources. This would require penalizing large values for $n$ using information criteria such as the Minimum Description Length (MDL), the Akaike Information Criterion (AIC)

The outputs of the GMCA are an array of dimension $(n_E,n)$ containing the spectral information of the components, and an array of dimension $(n,n_x,n_y)$ containing the spatial information of the components. In order to obtain $n$ normalized cubes of dimension $(E,x,y)$ we multiply each spectrum by its associated image. By collapsing these cubes along the $E$ axis, images of the retrieved sources can be obtained, and by collapsing them along the $x$ and $y$ axes we can obtain their spectra. The spectra can subsequently be used in \texttt{Xspec} or a similar analysis tool in order to fit physical models and retrieve physical parameters (see Section \ref{spectral}).

\section{Method performance}
\label{toy}

\subsection{Toy model definition}

%\begin{figure}
%\subfloat{\includegraphics[width = 9cm]{images/pearson_ratios_1Ms_arrows.pdf}}\\
%\subfloat{\includegraphics[width = 9cm]{images/pearsonr_ratios_100ks_comp_.pdf}}
%\caption{A comparison between the Pearson correlation coefficients of the input and output images of the Fe structure found for different intensity ratios Fe/Synchrotron by a GMCA, a GMCA with fixed Aref and an integration between $6.1$ keV and $7.1$ keV. On top, the GMCA is applied between $6.1$ keV and $7.1$ keV and the intensity corresponds to a $1$ Ms observation. On the bottom, the GMCA is applied between $5$ keV and $8$ keV and the intensity corresponds to a $100$ ks observation.}
%\label{fig:pearson_ratios}
%\end{figure}

\begin{figure*}[ht!]
  \subfloat{\includegraphics[width = 9.3cm]{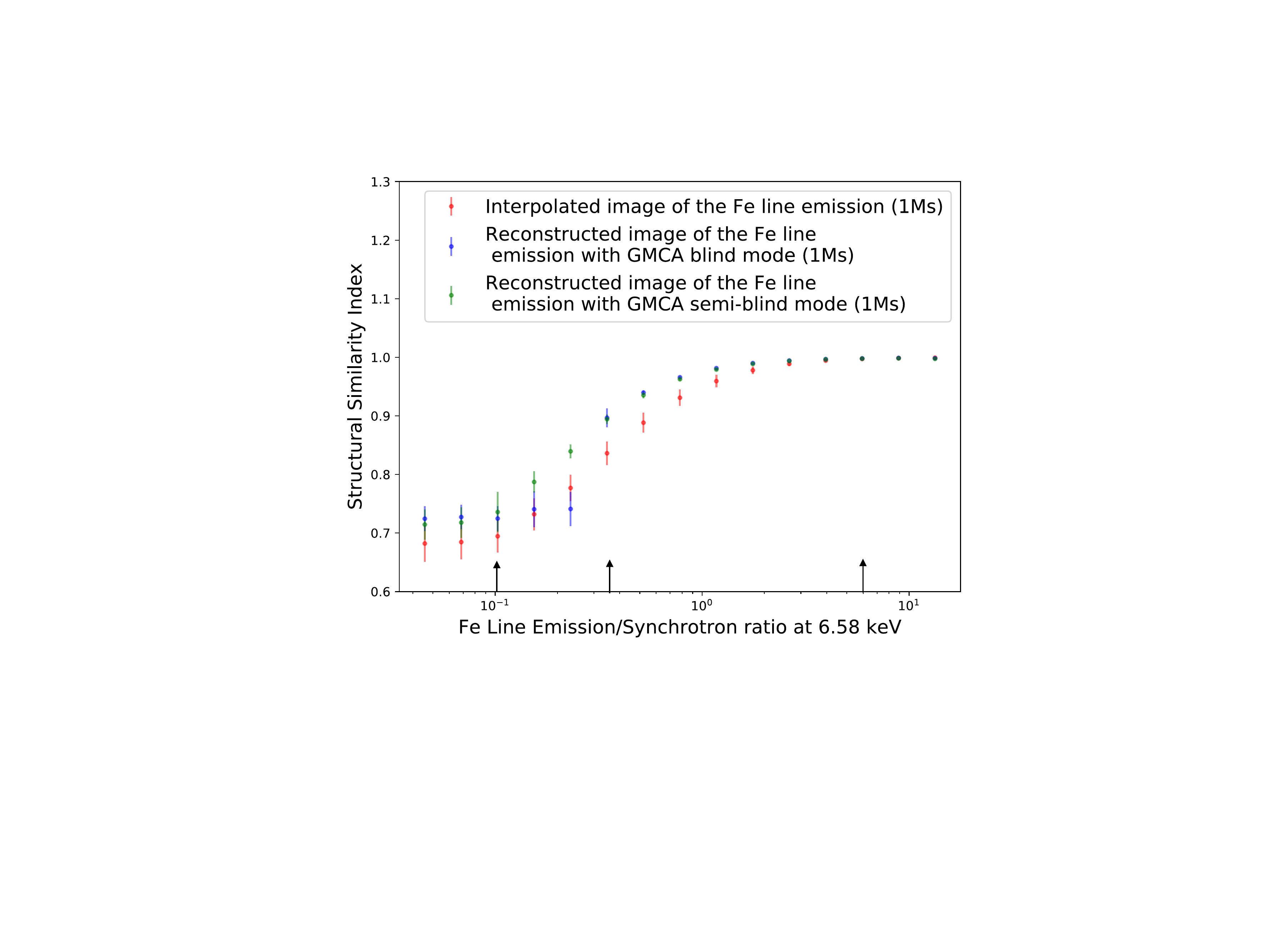}}
\subfloat{\includegraphics[width = 9.3cm]{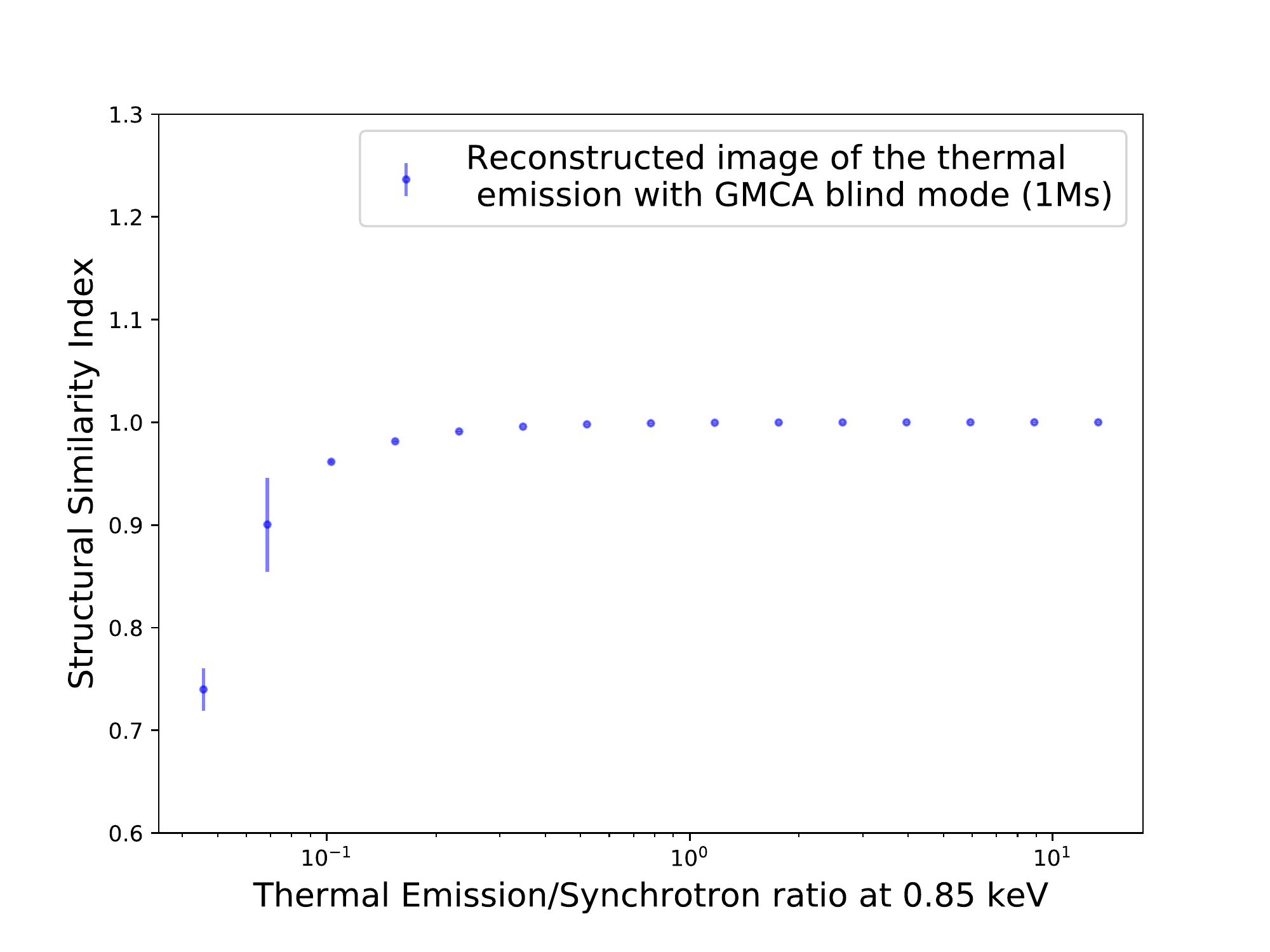}}
\vspace{-0.4cm}
\caption{SSIM coefficients of the input and output images found by GMCA for a total number of counts corresponding to a $1$ Ms observation. The points are the average of all Monte-Carlo realizations at a particular ratio, and the error bars are the standard deviation of those realizations. The left panel shows a comparison of the image quality obtained in retrieving the Fe structure in our first toy model for different line emission-to-synchrotron ratios, between an interpolation method, a GMCA in blind mode, and a GMCA in semi-blind mode. Some images corresponding to the ratios indicated by arrows are shown in Figure \ref{fig:ssim_examples} The right panel shows the image quality of the thermal emission structure retrieved for different ratios by a GMCA in blind mode.}
\label{fig:ssim_ratios}
\end{figure*}

        To test the performance of the GMCA in disentangling components in X-ray data, we designed toy models inspired by real X-ray observations of SNRs. We chose to simulate a SNR similar to Cassiopeia A, one of the best-studied SNRs and one which has benefited from deep megasecond observations. 

        Our toy models consist of a data cube composed of the sum of individual components to which we add Poisson noise. Each component comprises an image multiplied by a spectrum (see Figure \ref{fig:toy_model}). The images were obtained by applying the GMCA to real {\it Chandra} data from Cassiopeia A (see Section \ref{sect:real data}), and smoothing the output to mitigate the noise. For now, the relevance of these images is not important: we only want to ascertain if, when the components are known, the GMCA is able to disentangle them when mixed together. The spectra we use are the theoretical spectra folded through the {\it Chandra} instrument response; the energy binning is $43.8$ eV (three times the native energy channel width), and the pixel size is $1.8$ arcsec (four times the native pixel size). We also add a completely flat image associated with the instrumental background\footnote{Derived from closed/stowed observations available at: \url{http://cxc.harvard.edu/ciao/download/caldb.html}} to better simulate observed data. We do not add a cosmic X-ray background, because this background being isotropic at the scale of CasA, its spatial template would be a flat image, and therefore the addition of a cosmic X-ray background and the instrumental background would only end up being one component, with a slightly different spectrum. Finally, we generate Poisson noise. In this study we begin by focusing on two typical observational scenarios (see Table \ref{table:Toy_model}): synchrotron continuum emission entangled with line emission (Model 1), and synchrotron continuum emission entangled with thermal emission (Model 2). In both models, we set the synchrotron emission as one with the highest total number of counts.

\begin{table}
\caption{Description of the toy models. For all models, $N_{\rm H}$ is set to $N_{\rm H}=0.5\times 10^{22}$ cm$^{-2}$. For the thermal model (\textit{apec)}, the ionization timescale is set to $\tau=1\times 10^{10}$cm$^{-3}$s and the abundances to solar values.}             % title of Table
\label{table:Toy_model}      % is used to refer this table in the text
\centering                          % used for centering table
\begin{tabular}{c | l | l}        % centered columns (4 columns)
\hline                 % inserts double horizontal lines   % table heading 
\hline
   & Description & Parameters \\
  \hline
  Model 1 & Power Law + & $\Gamma=2.0$ \\
   & Gaussian & $E_c=6.58$ keV \\ & & $\sigma=80$ eV \\
   \hline
 Model 2 & Power Law + & $\Gamma=2.0$ \\
   & Apec  & $kT=2$ keV \\
   \hline
 Model 3 & Power Law + & $\Gamma=2.0$ \\
   & Two Gaussians  & $E_{c1}=6.55$ keV \\
   &  & $E_{c2}=6.64$ keV \\ & & $\sigma=0$ eV \\
   \hline
 Model 4 & Power Law + & $\Gamma=2.0$ \\
   & Two Apecs  & $kT_1=2$ keV \\
   & & $kT_2=0.5$ keV \\
                                 %inserts single line
\end{tabular}
\end{table}        
 
\begin{figure*}[ht!]
  \subfloat{\includegraphics[width = 6cm]{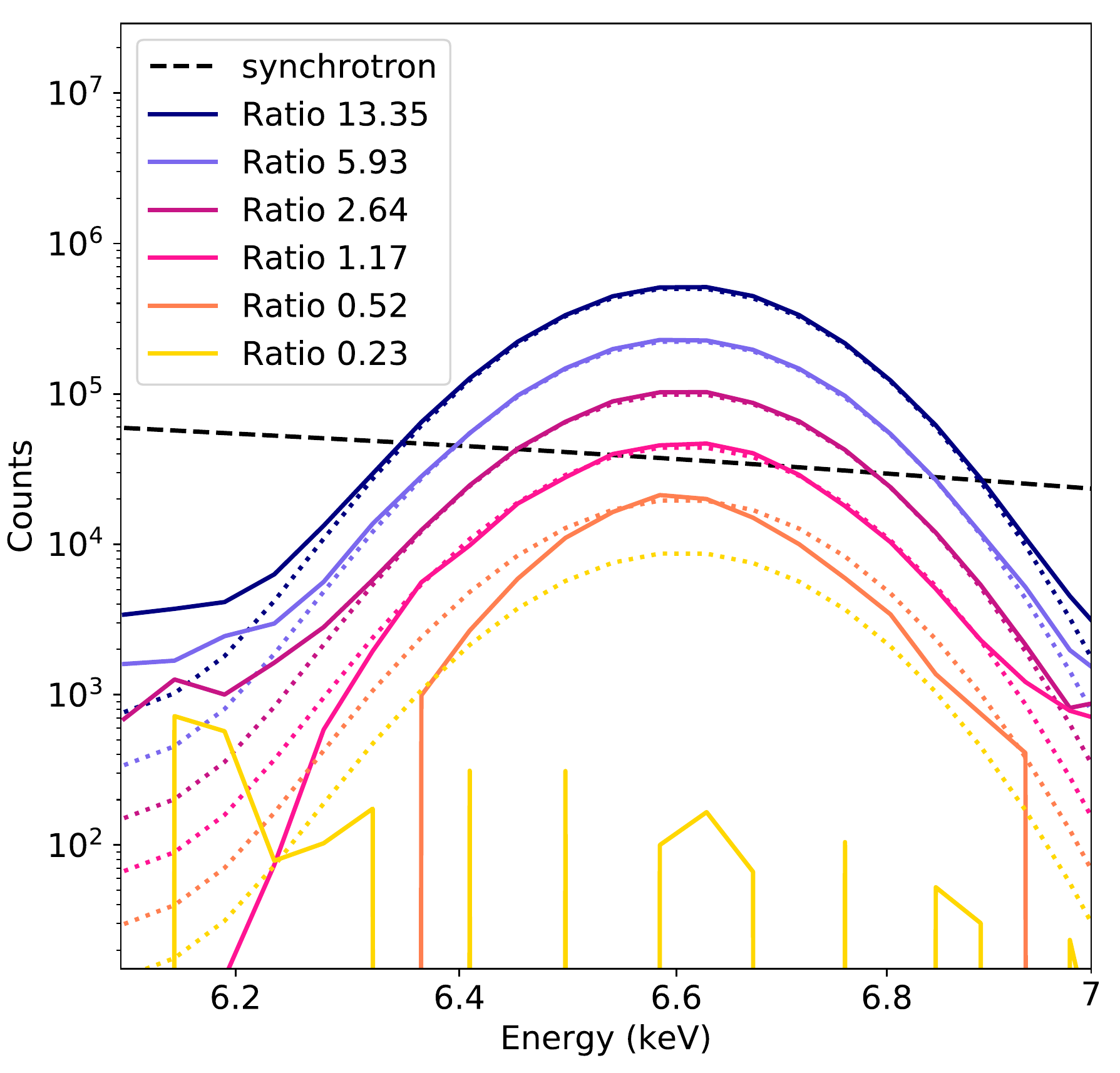}}
\subfloat{\includegraphics[width = 6.25cm]{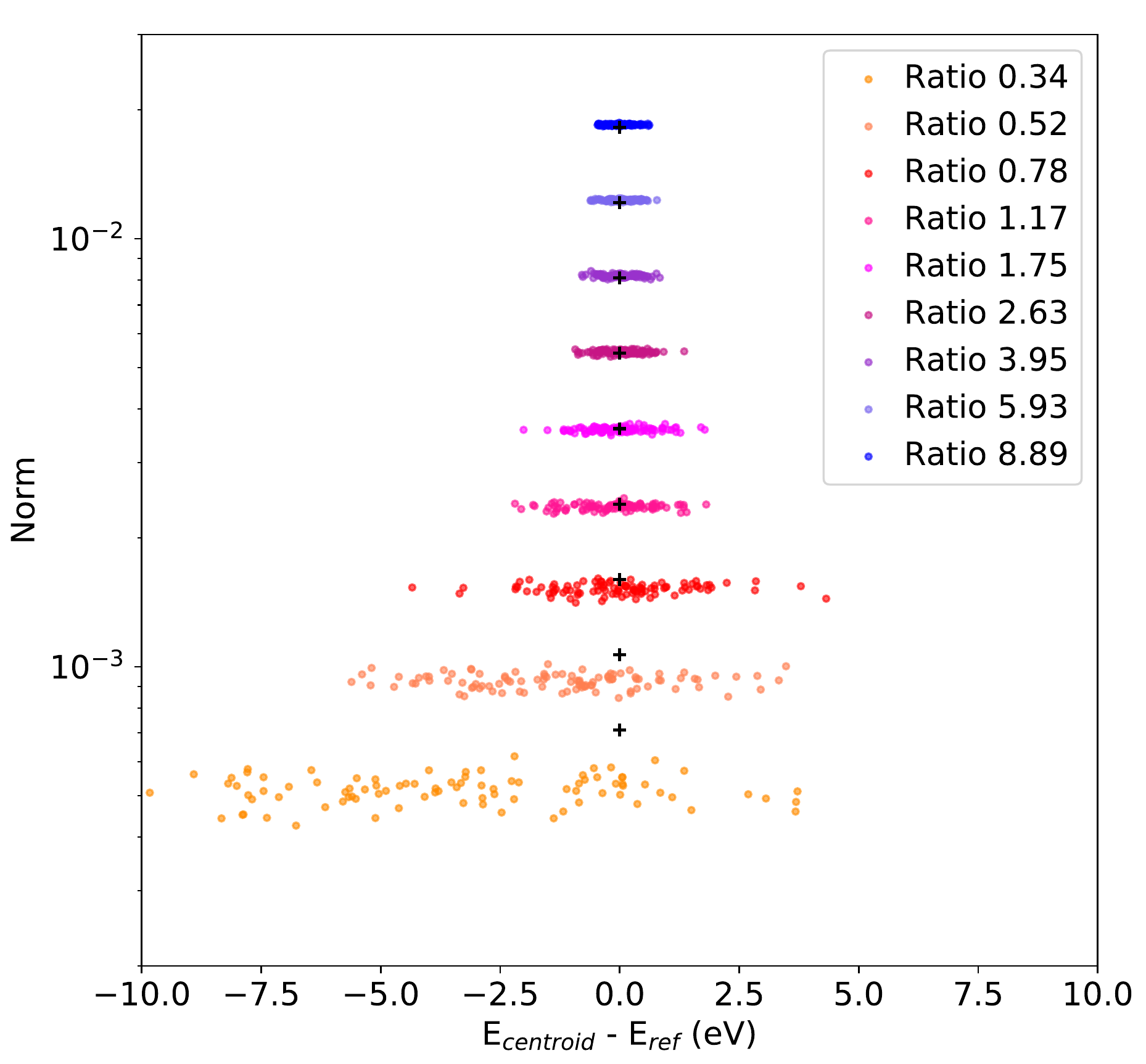}}
\subfloat{\includegraphics[width = 6.15cm]{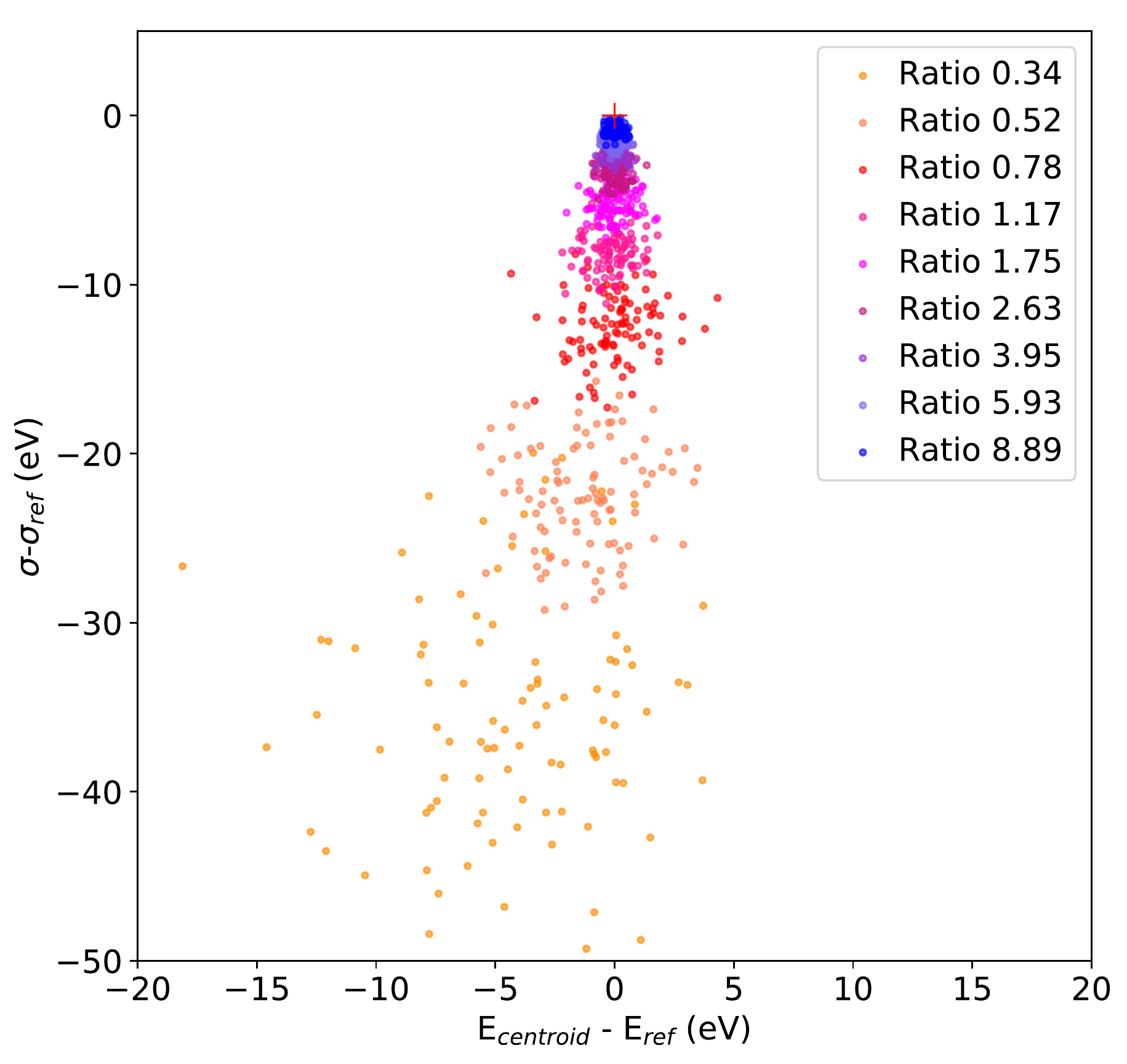}}
\caption{Left: Spectra retrieved by GMCA for different Fe line-to-synchrotron ratios in our first toy model with a total number of counts corresponding to a $1$ Ms observation. Retrieved spectra are shown as solid lines,  theoretical spectra as dotted lines. The synchrotron spectrum is displayed as an indication of the relative strengths. The two other plots represent parameters of the Fe K component retrieved by GMCA for 100 realizations of each of nine different Fe line-to-synchrotron
ratios with a total number of counts corresponding to a $1$ Ms observation. Center: Retrieved $E_c$ and norm. Right: Retrieved $E_c$ and $\sigma$.  In both cases, the theoretical results are represented by crosses.}
\label{fig:model1}
\end{figure*}

\begin{table}
\caption{Equivalence between the max(Fe or thermal emission)-to-synchrotron ratios and the physically more significant flux ratios for our four toy models. The components are named after their main characteristic ($E_c$ for the Gaussians, $kT$ for the thermal emissions), as they are listed in Table \ref{table:Toy_model}. The energy ranges listed below each component are those from which the ratios are calculated.}             % title of Table
\label{table:ratios}      % is used to refer this table in the text
\centering                          % used for centering table
\begin{tabular}{c || c | c | c | c  }        % centered columns (4 columns)
\hline                 % inserts double horizontal lines   % table heading 
\hline
Models & Model 1 & Model 2 &Model 3& {Model 4} \\
\hline 
 Comp.&$E_{c}$&$kT$&$E_{c1}$&$kT_2$\\
 \hline
 \hline
 &$6.2$ -&$0.5$ -&$6.4$ -&$0.5$ -\\
Ratios&$7$ keV&$8$ keV&$7$ keV&$4$ keV\\
  \hline 
13.35 & 4.20 & 2.39 & 4.36 & 1.67\\
  \hline
8.90 & 2.80 & 1.59 & 2.91 & 1.11\\
  \hline
5.93 & 1.86 & 1.06 & 1.93 & 0.74\\  
  \hline
3.95 & 1.24 & 0.71 & 1.29& 0.50\\  
  \hline
2.64 & 0.83 & 0.47 & 0.86 & 0.33\\  
  \hline
1.76 & 0.55 & 0.31 & 0.57 & 0.22\\  
  \hline
1.17 & 0.37 & 0.21 & 0.38 & 0.15\\  
  \hline
0.78 & 0.25 & 0.14 & 0.26 & 0.098\\  
  \hline
0.52 & 0.16 & 0.093 & 0.17 & 0.065\\  
  \hline
0.35 & 0.11 & 0.062 & 0.11 & 0.043\\  
  \hline
0.23 & 0.073 & 0.041 & 0.076 & 0.029\\  
  \hline
0.15 & 0.049 & 0.028 & 0.050 & 0.019\\  
  \hline
0.10 & 0.032 & 0.018 & 0.033 & 0.013\\  
  \hline
0.069 & 0.0022 & 0.012 & 0.022 & 0.0086\\ 
  \hline
0.046 & 0.0014 & 0.0082 & 0.015 & 0.0057\\  
%inserts single line
\end{tabular}
\end{table}       

        The results of our method depend on the relative level of the Poisson noise, and therefore on the total number of counts in the signal. This parameter is chosen in order to reflect the reality of the data we get from spectro-imaging instruments. Hence we set the count rate of the synchrotron and line or thermal emission to be of the order of that observed in Cassiopeia A. We then simulated two datasets, corresponding to a $1$ Ms or a $100$ ks observation with the {\it Chandra} ACIS-S instrument.
    
        The ratio between the strength of the main component and that of the secondary components is also an essential factor. For Model 1, we define this as the Fe line-to-synchrotron ratio at $6.58$ keV (the peak of the Gaussian); for Model 2, it is defined as the thermal emission-to-synchrotron ratio at $0.85$ keV. We progressively decrease the contrast of the second component relative to that of the synchrotron emission following $15$ ratios.
    
    For both toy models we tested the same ratios. Table \ref{table:ratios} presents a conversion table between these ratios and the Fe line-to-synchrotron flux ratios  between $6.2$ and $7$ keV, or the thermal emission-to-synchrotron ratio in the $0.5-8$ keV band.

\begin{figure*}[ht!]
\subfloat{\includegraphics[width = 9.3cm]{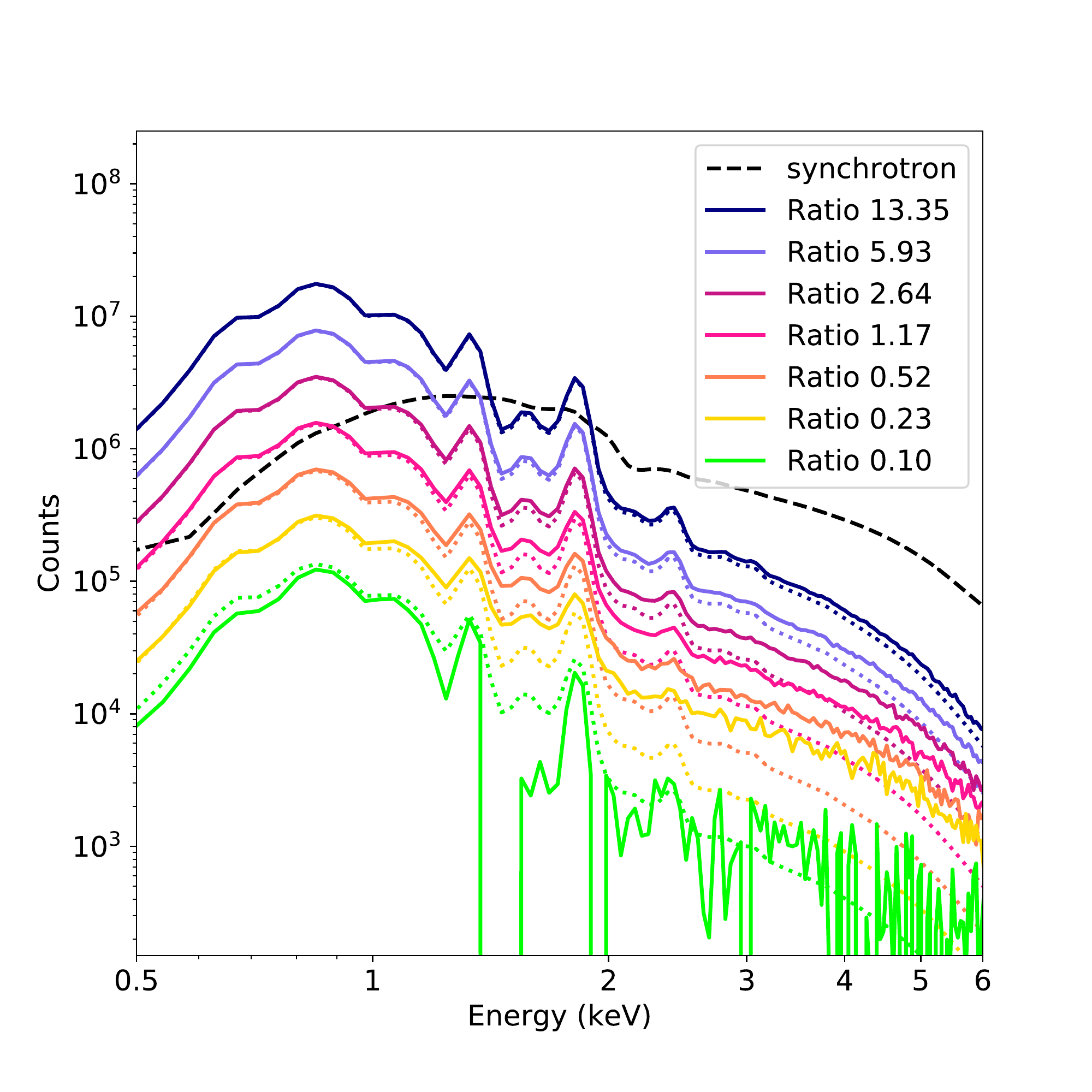}}
{\includegraphics[width = 9.cm]{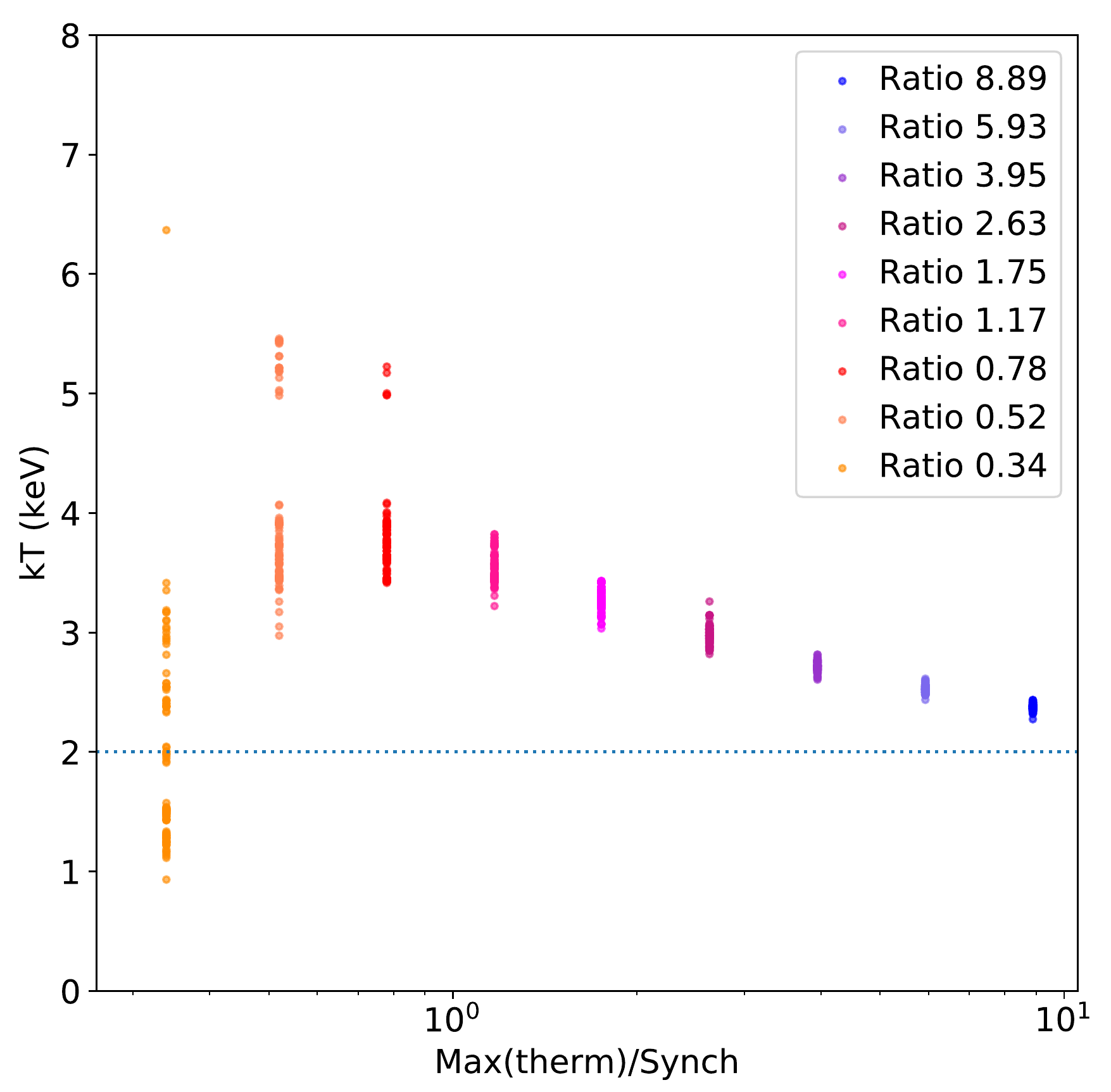}}
\vspace{-0.2cm}
\caption{Left: Spectra retrieved by GMCA for different thermal emission-to-synchrotron ratios with a total number of counts corresponding to a $1$ Ms observation. Retrieved and theoretical spectra are shown in solid and dotted lines, respectively. The synchrotron spectrum is displayed as an indication of the relative strengths. Right: kT retrieved by \texttt{Xspec} by fitting a thermal model on the thermal emission spectra retrieved by GMCA for different Fe line-to-synchrotron ratios with a total number of counts corresponding to a $100$ ks observation. For each ratio, we made a Monte-Carlo with 100 realizations, fitting in \texttt{Xspec} a thermal model on every realization associated with the error bars given by the Monte-Carlo. The theoretical kT is $2$ keV and is indicated by the blue dotted line.}
\label{fig:model2}
\end{figure*}    
 
\subsection{Reconstructed image fidelity}
\label{imagefidelity}

        To assess the accuracy of the results of the GMCA, we compared both the similarities between the input and the output images, and the reliability of the spectral parameters fitted. For the image benchmarks, we used the structural similarity index \citep[SSIM; see][]{zhou04}, which measures the perceived similarities between two images by incorporating perceptual phenomena and the idea that close pixels have strong interdependencies, instead of solely measuring absolute differences. This index takes the form of a number between zero and one, one being a perfect resemblance and zero indicating perfect dissimilarity. In our case, below an SSIM of $0.75$ we can consider that the source has not been retrieved, the remaining correlations being linked to the similarities between the synchrotron image, the Fe image, and the Poisson noise associated to them.
    
    For each line-to-synchrotron ratio we then performed a Monte-Carlo simulation of 100 different Poisson realizations to test the robustness of the algorithm. We compared the results of the GMCA in pure blind mode with that of the GMCA in semi-blind mode, with the theoretical shape of the Fe line fixed. We also compared these results to that of an interpolation method between $6.1$ and $7.1$ keV (the left panels of Figs. \ref{fig:ssim_ratios} and \ref{fig:ssim_ratios2} show the results for the simulated $1$Ms and 100 ks observations, respectively). This method consists of estimating the underlying synchrotron spectrum between $6.1$ and $7.1$ keV by interpolating it. The synchrotron image is then determined by integration (e.g., between $5$ and $6$ keV, where the Fe is absent) and the synchrotron cube is obtained by multiplying this image and the interpolated spectrum. Subsequently, we subtract the aforementioned cube, and the synchrotron-subtracted remaining cube constitutes an estimation of the Fe structure.

        For both simulated exposures, we see that Fe line-to-synchrotron ratio images given by the GMCA have slightly better SSIM coefficients to those obtained with an interpolation method. However, a sudden drop in the GMCA results points out the moment when the algorithm in blind mode is no longer able to find the Fe structures. The descent is smoother with a fixed spectrum (semi-blind mode) because the algorithm is given more information to search for potential sources, but as the number of counts in the iron line decreases the noise increases. In blind mode, the GMCA retrieves an image of the Fe spatial structure when it is up to $2.9$ times weaker than the synchrotron in the case of a total number of counts corresponding to a $1$ Ms observation ($9$ times weaker in flux), and up to $1.8$ times higher than the synchrotron for $100$ ks ($1.8$ times weaker in flux).
    
    The GMCA in blind mode does not benefit from the information that the Fe line is contained between $6.1$ and $7.1$ keV, but still gives very good results. Furthermore, the interpolation method cannot be used on components whose spectra are extended on an energy range that is too wide, as we see with our second toy model. Also, we see in Sect. \ref{sect:real data} that with real data, what looks like a Gaussian can contain some hidden information that a GMCA in blind mode will be able to retrieve, but an interpolation can only find the Gaussian as a whole.
    
    The fact that the GMCA gives good images until it is suddenly unable to find anything but noise suggests that the algorithm can be trusted; in this particular test the Fe distribution is found or is not, but the algorithm never gives images of spurious, over-interpreted structures (see \ref{fig:ssim_examples} for an example of images becoming noisier as the component becomes fainter). In our test case, when we increase the number of sources, the first two remain the synchrotron and Fe structure, the rest are only noise. As our data are Poissonian, the noise component has a shape similar to that of the main component, here the synchrotron, with large fluctuations.
    
    We proceeded in the same way with our second toy model, featuring a synchrotron continuum emission and a thermal emission (see the 
right panel of Figs. \ref{fig:ssim_ratios} and C.2 for the simulated $1$Ms observation and  the $100$ ks one, respectively). Here, the comparison with an interpolation method is impossible because the thermal spectrum cannot be subtracted from the synchrotron with a simple interpolation. The GMCA in semi-blind mode does not make sense either, because with real data it would be impossible to know the shape of a thermal emission a priori. With a total number of counts corresponding to a $1$ Ms observation, the GMCA in blind mode applied from $0.5$ to $10$ keV is able to retrieve an image of the thermal emission spatial structure when this component is up to $14.6$ times weaker than the synchrotron ($83.3$ times weaker in flux). With a total number of counts corresponding to a $100$ ks observation, it could retrieve an image up to $4.3$ times weaker than the synchrotron ($13.7$ times weaker in flux). The thermal emission in our second toy model can be retrieved with smaller ratios than the Fe line because it is non-negligible on a wider energy range, providing more counts to the algorithm.
    
    We note that the instrumental background was not retrieved in any of the cases, and that it did not leak on any other component. This is due to the fact that the two largest wavelet scales being eliminated, the instrumental background associated with a flat image, were automatically suppressed with it.
    
\subsection{Spectral fidelity}
\label{spectral}

\begin{figure*}[ht!]
\subfloat{\includegraphics[width = 8.9cm]{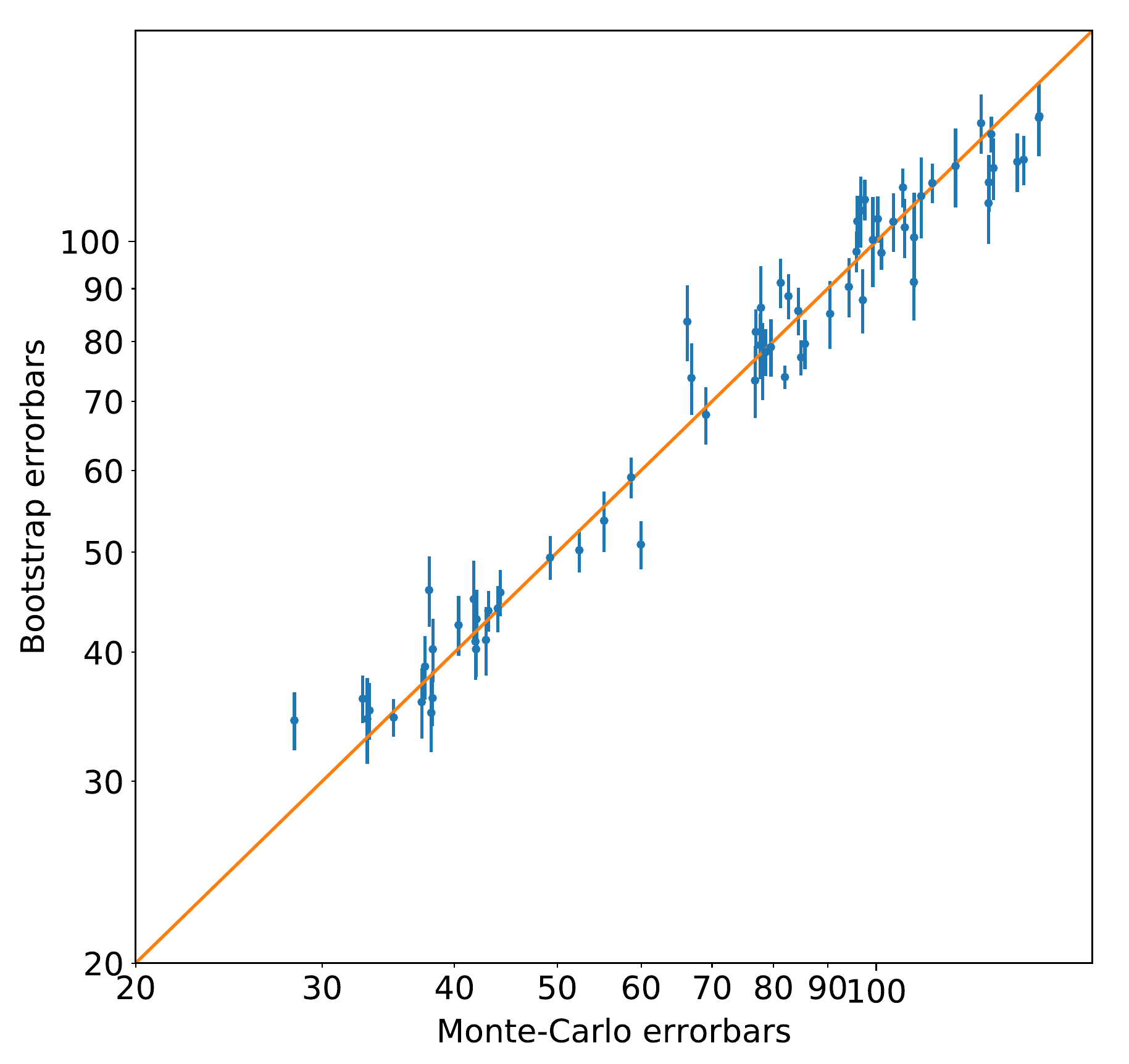}}
\subfloat{\includegraphics[width = 9.2cm]{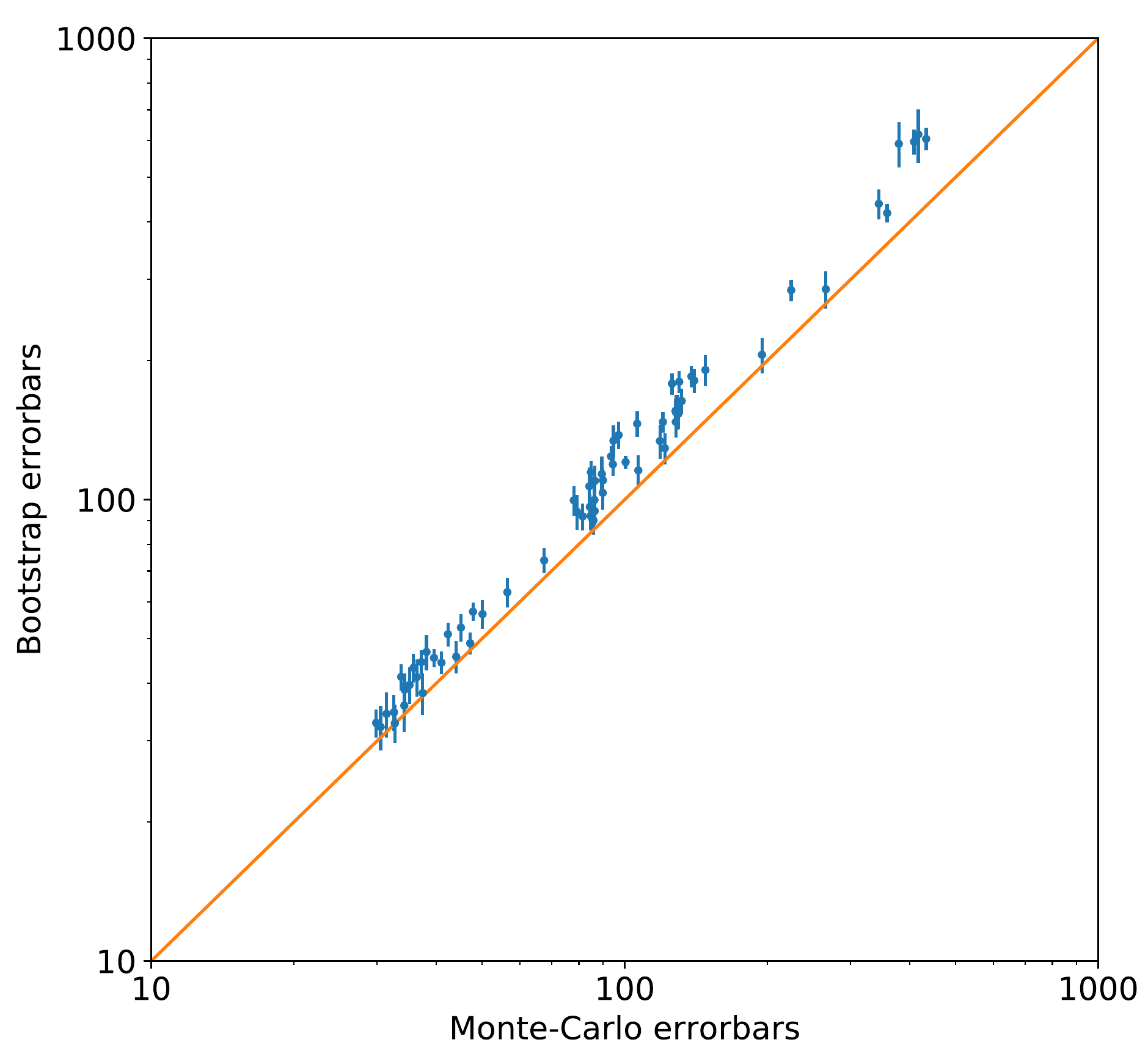}}\\
\vspace{-0.5cm}
\caption{Correlation between the error bars given by the standard deviation of 100 Monte-Carlo realizations and the error bars given by the standard deviation of 100 simulated realizations resampled out of a single one thanks to the block bootstrap method, with and without GMCA. Left: Spectra are those of the Fe Gaussian in our first toy model with a total number of counts corresponding to a $100$ ks observation and a line-to-continuum ratio of $5.93$ (in the absence of a synchrotron spectrum, this ratio characterizes the  norm of the Gaussian and the ratio of $5.93$ corresponds to the fourth norm in our previous tests). Right: Monte-Carlo vs. Block bootstrap error bars on the results given by GMCA on our first toy model, for the same duration and ratio. In both cases, we made 100 Monte-Carlo realizations and we used a block bootstrap resampling on ten realizations, 100 times each, for a block length of $78$ (cube root of the total number of events) in order to evaluate the influence of the initial realization on a block bootstrap resampling.}
\label{fig:errorbars}
\end{figure*}

        For every Fe line-to-synchrotron ratio for which the Fe K distribution is found by the GMCA in blind mode, the retrieved spectrum is comparable to the input spectrum with some noise appearing as the Fe component becomes fainter (see left panel of Fig. \ref{fig:model1}). Apart from a slight overestimation of the wings, the retrieved spectra are accurate and their normalizations well estimated. 
    
    In order to obtain a  more precise estimate of the spectral accuracy of the method, we fitted the recovered spectra in \texttt{Xspec} and compared the parameters thus obtained with a fit of the original data without GMCA processing directly in \texttt{Xspec}. Fitting the retrieved spectra requires estimating the errors for every spectral bin. In spite of the fact that our input data are Poissonian, we cannot assume that the results given by the GMCA will still be such. Therefore, we used the standard deviation of 100 Monte-Carlo realizations as an estimation of the error.
    
    We tested the accuracy of the spectra retrieved by the GMCA in Model 1 by comparing their centroids, widths, and normalizations to the theoretical ones  (see central and right panels of Fig. \ref{fig:model1}). The norms are almost perfectly retrieved (left and central panels of Fig. \ref{fig:model1},), and even the slight energy shift for the smaller ones (around $5$ eV) is negligible as compared to the instrument resolution, which is $150$ eV (at $5.9$ keV) for the ACIS-S camera \footnote{ \url{http://cxc.harvard.edu/cal/Acis/} }. The wings are a little overestimated in the first norms (Figure \ref{fig:model1}, left panel), while the width $\sigma$ is underestimated in the last ones (left and right panels of Fig. \ref{fig:model1}). It may be due to the fact that in the fainter part of the Gaussian, the signal is largely dominated by the synchrotron, which makes the disentanglement harder than at the peak of the Gaussian.
    
    We made the same comparison with the Gaussians recovered without using GMCA by fitting a power law and a Gaussian on the original spectra in \texttt{Xspec} (see Fig. \ref{fig:spectra_noGMCA}). The retrieved norms and centroids are a little more accurate (Fig. \ref{fig:spectra_noGMCA}, left panel), but are relatively similar to the results given after GMCA. However, the retrieved $\sigma$ are not underestimated, and are still centered on the theoretical value for low ratios (Fig. \ref{fig:spectra_noGMCA}, right panel). Thus, the GMCA introduces a bias in calculating some physical parameters in \texttt{Xspec}, but this bias is minimal compared to the $150$ eV spectral resolution.

    Finally, we tested the accuracy of the spectra retrieved by GMCA in our second toy model, featuring a synchrotron and a thermal emission (see left panel of Fig. \ref{fig:model2}). The spectra are mainly well retrieved, even for low thermal emission-to-synchrotron ratios, but they are always overestimated at high energies. This reflects the fact that the synchrotron is contaminating the thermal emission: because of the spatial overlap between the two structures, there is a leakage from the main one into the weaker one when the number of counts is too low. This leakage strongly impacts the temperature retrieved after a fitting in \texttt{Xspec}, the necessary information being the slope at high energies. As shown in the  right panel of Fig. \ref{fig:model2}, the overestimation of the spectra, greater as the ratio decreases, is directly affecting the retrieved $kT$. However, $kT$ is a global parameter, relying on the information contained over the full energy range, thus highly susceptible to being impacted by an overestimation at high energies. Local parameters, like $N_{\rm H}$ or $\tau$, are almost perfectly estimated for thermal-to-synchrotron ratios as low as $0.52$. For example, the theoretical $N_{\rm H}$ is equal to $0.5\times 10^{22}$ cm$^{-2}$, and for a ratio of $3.95$ we retrieve $(0.490 \pm 0.001)\times 10^{22}$ cm$^{-2}~$  and for a ratio of $0.34$, we obtain $(0.485 \pm 0.008)\times 10^{22}$ cm$^{-2}$ where errors are the standard deviation on 100 Monte-Carlo realizations. In the same way, the theoretical $\tau$ is $1\times 10^{10}$ cm$^{-3}$~s; for a ratio of $3.95$ we retrieve $(9.13 \pm 0.05)\times 10^{9}$ cm$^{-3}$~s, and for a ratio of $0.34$, we obtain $(9.12 \pm 0.27)\times 10^{9}$ cm$^{-3}$~s.

\section{Implementing a new inpainting step in the GMCA}
\label{inpainting}

    In this section we discuss the introduction of an extra step in the GMCA algorithm based on an inpainting method. Inpainting is a process consisting in reconstructing parts of an image that are lost or willingly removed. In photography, it can be used to clean the image, removing defaults or inappropriate details. This tool was shown to be useful to improve our blind source separation method.
    
        We previously saw that in the results given by the GMCA on toy models composed of two physical sources there could be some leakage from the main component to the other one (e.g., leakage of the synchrotron component to the thermal component in Fig. \ref{fig:model2}). These leakages are often balanced by negative parts in the image or spectrum of the main component. In order to correct that leakage, we added an extra step to the GMCA.
    
    The GMCA being an iterative algorithm, our revised version retains a loop of about $150$ iterations of the usual algorithm, followed by a smaller loop with a new step in which each result of the previous state is treated in a way to forbid negative values. To do so, a first method would be to define a mask where the reconstructed images take negative values, and apply those masks to the wavelet transforms of those images, $S$, imposing a  zero value on the negative parts before they are processed to estimate $A$. The results can be improved by replacing the raw masking by an inpainting, here a reconstruction of the masked parts of the image using a wavelet transform \citep[see][]{fadili2007}. We do this in order to constrain the algorithm to converge to a more physical solution.
    
    Our new loop can be described thus:

\begin{itemize}
\item Step 1 : Estimation of $S$ thanks to $X$ and the previous $A$.
\item Step 2 : Defining masks set to zero where the reconstructed images are negative, indicating an area where strongly correlated components are overlapping, and one elsewhere.
\item Step 3 : Inpainting of $S$ (in wavelets) using the previously defined masks.
\item Step 4 : Estimation of $A$ for fixed $S$ by minimizing $\| X-AS \|_F$.
\end{itemize}

\begin{figure*}[ht!]
{\includegraphics[width = 9.3cm]{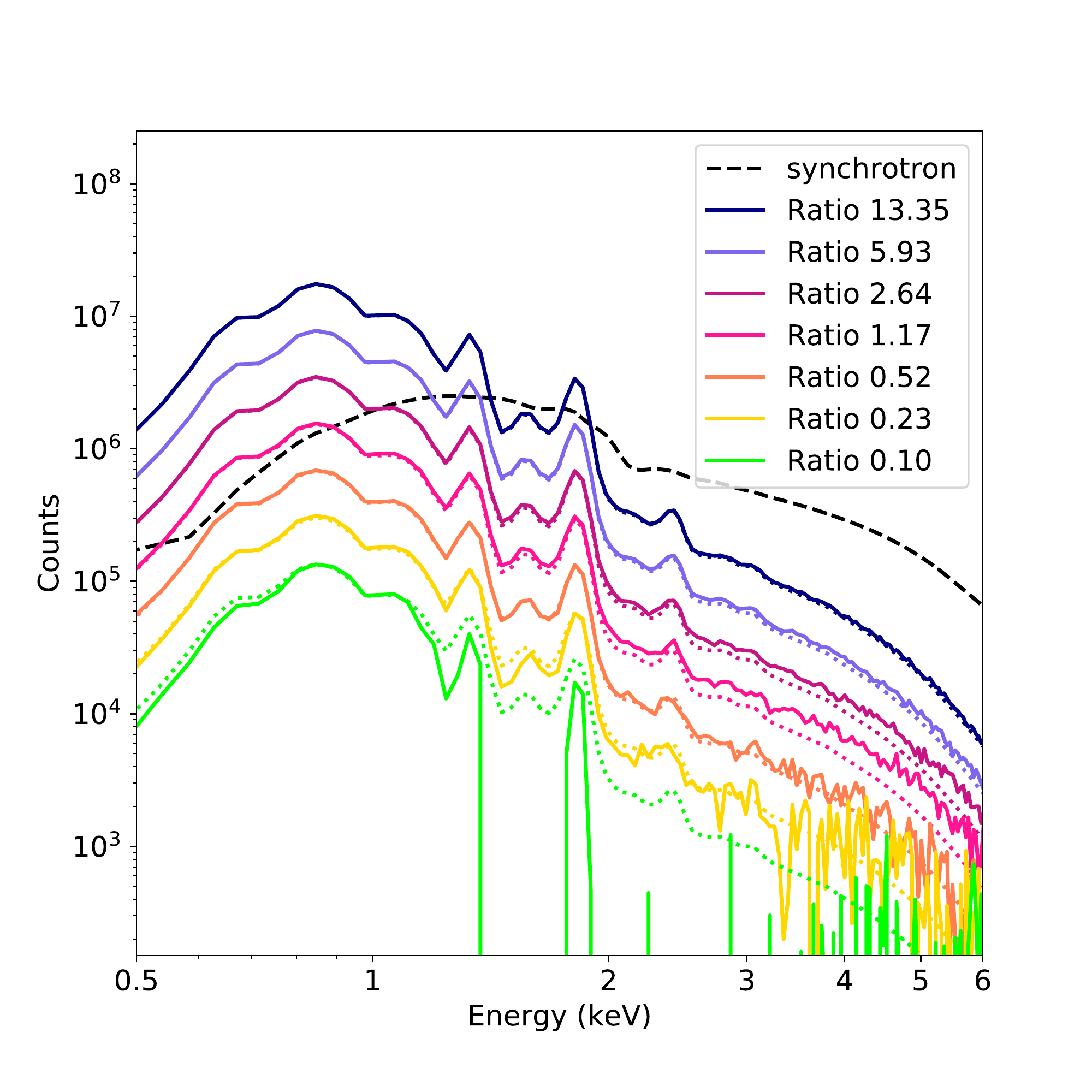}}
{\includegraphics[width = 9.05cm]{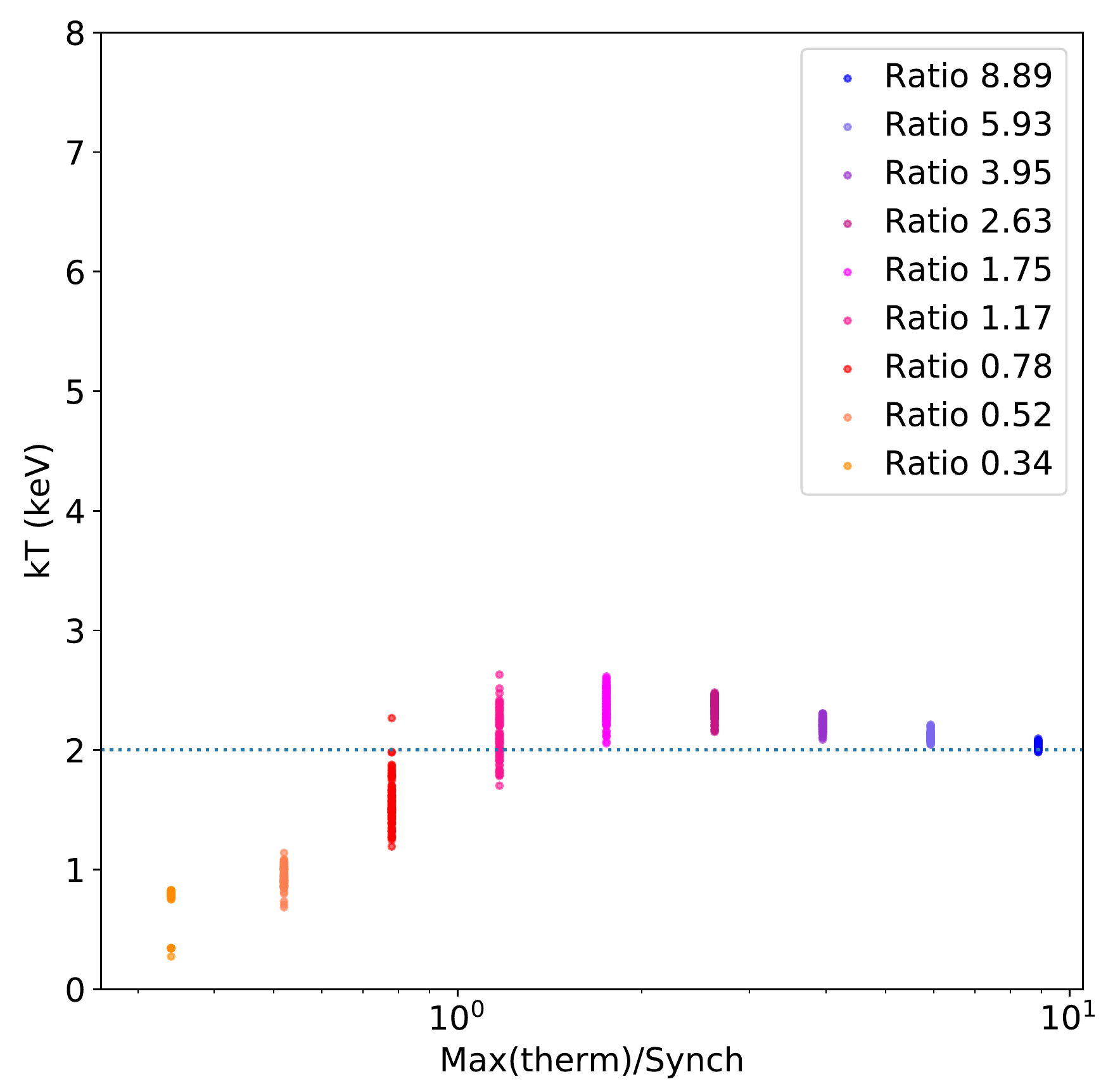}}
\vspace{-0.5cm}
\caption{As in Fig. \ref{fig:model2} after the inpainting step described in Sect. \ref{inpainting}.}
\label{fig:inpainting_results}
\end{figure*}

    As can be seen in Fig. \ref{fig:inpainting_results}, our inpainting step accurately corrects the leakage from the synchrotron to the thermal emission component in our second toy model: the retrieved spectra are closer to the truth. The resulting impact on the fitting in \texttt{Xspec} is also significant, as the temperatures are now more accurately retrieved for sufficiently high thermal emission-to-synchrotron ratios (see Fig. \ref{fig:inpainting_results}). The convergence of our new loop is not mathematically proven, but we empirically noted that the solution stabilized quickly. In the science cases that we explored, three iterations were sufficient to recover more accurate spectral results.

\section{Estimating errors with only one realization}
\label{errors}

    The Monte-Carlo method cannot be used to retrieve error bars with real data, as only one observation is available: the observed one. Therefore, a resampling method such as the Bootstrap \citep[see][]{befron79}, able to simulate several realizations out of a single one, is necessary.

\subsection{Block bootstrap}
\label{sec:boot}
        The bootstrap is a statistical method consisting of a random sampling with replacement from a current set of data. If the initial data is a collection of $N$ events, a resampling obtained through bootstrapping would be a set of $N$ events taken randomly with replacement amid the initial ones. This method can be repeated in order to simulate as many realizations as needed to estimate standard errors or confidence intervals. In order to save calculation time, we choose to resample blocks of data of a fixed size instead of single events: this method is named block bootstrap.
    
    In our case, the data is the set of all photons detected by an X-ray telescope during its observation time, each photon being considered as a triplet $(E,x,y)$. Because of the massive amount of events, we use a block bootstrap resampling method. The ordering variable is time, independent of $(E,x,y)$, and therefore defining blocks preserves the random character. There is no proper way to choose a block length a priori; a few tests seem to indicate that a length of the order of the cube root of the total data set size is efficient with our type of data.
    
    The errors on the spectra are calculated as the standard deviation of the values on each energy bin over all new samples. The error on the i-th bin is thus:
    
\begin{equation}
error[i]=\sqrt[]{\frac{\sum\limits_{j=1}^n \bigl( spec[i,j]-\overline{spec[i]}\bigr)^2 }{n}}
,\end{equation}

where $n$ is the number of resamples, $spec[i,j]$ the value of the spectrum in the i-th bin of the j-th sample, and $\overline{spec[i]}$ the mean of the values of the spectra in the i-th bin over the $n$ resamples.

\subsection{Estimated errors}

        Our aim in using a block bootstrap resampling method is to estimate errors on the spectral data points that will allow us to fit spectra issued from real data in \texttt{Xspec}. In the first place, we compared the error bars given by 100 Monte-Carlo realizations of the Fe Gaussian alone to those retrieved by these methods out of a single one. The data we used were the Fe Gaussian of our first toy model between $5$ and $8$ keV for a $100$ ks observation and a ratio of $5.93$. 
    
    To do so, for every energy bin we looked at the correlation between the standard deviation of the spectral values as given by a Monte-Carlo and by the block bootstrap method; by applying the resampling method to different realizations we were able to evaluate the errors on the bootstrap error bars (i.e., the uncertainty induced by using one given observation). In Fig. \ref{fig:errorbars}, we see that the error bars obtained through resampling are consistent with the Monte-Carlo error bars.

    To find out if applying the GMCA algorithm introduces a bias, we also compared the error bars given by the standard deviation of 100 GMCA applied on different Monte-Carlo realizations and the error bars given by 100 GMCA applied on 100 resamples.
    
\begin{figure*}[ht!]
\subfloat{\includegraphics[width = 9.2cm]{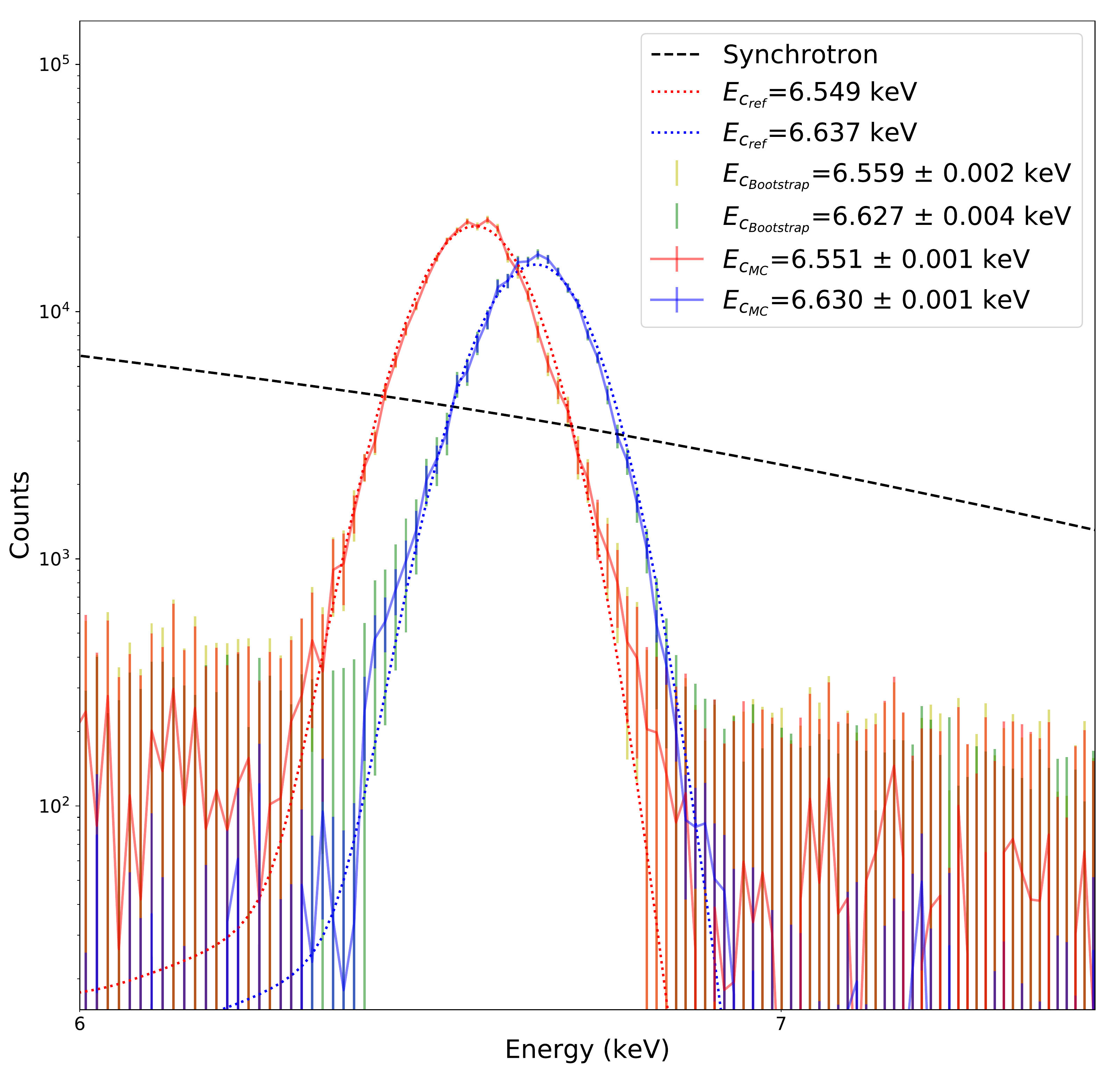}}
\subfloat{\includegraphics[width = 9.2cm]{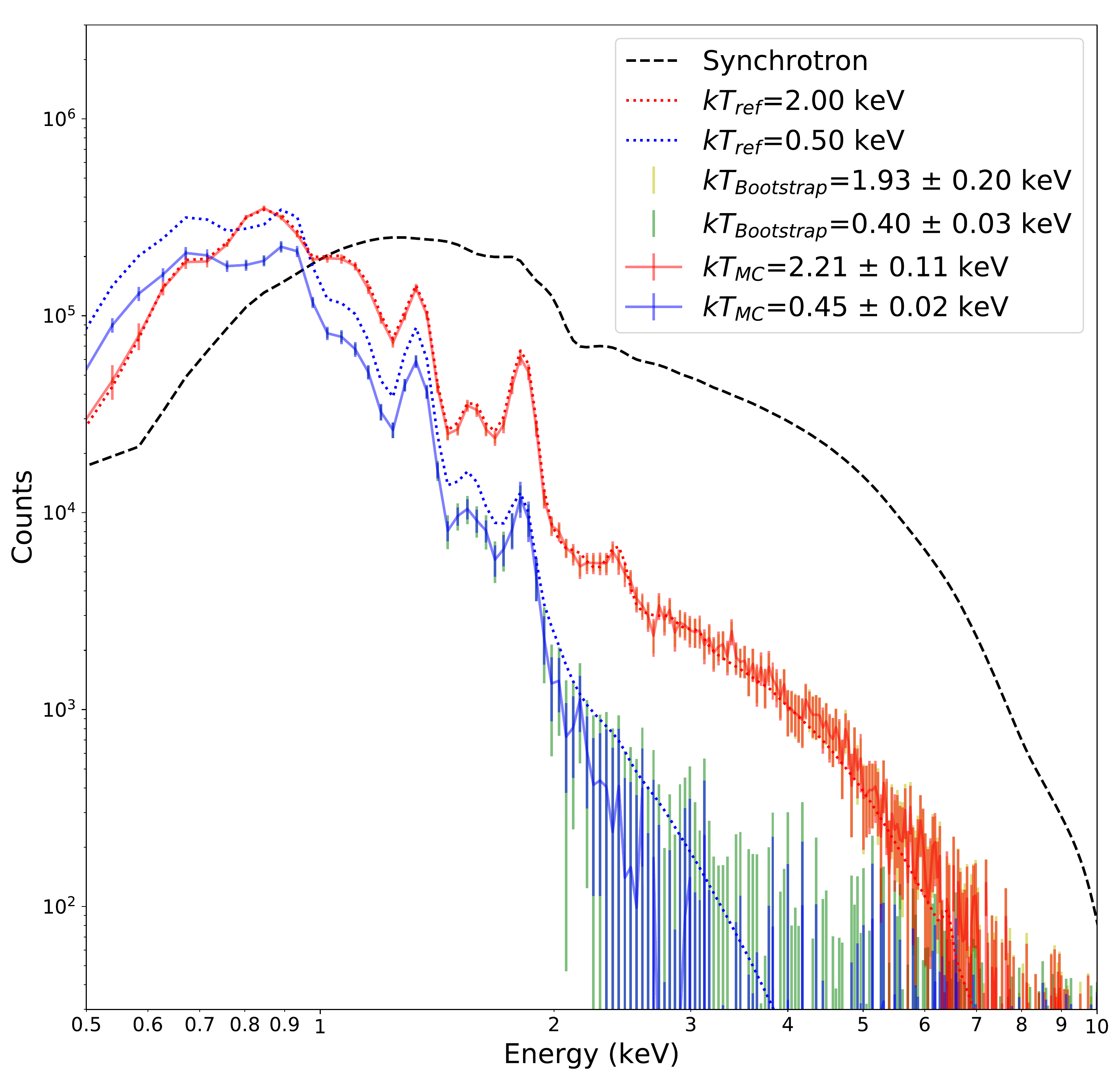}}
\vspace{-0.25cm}
\caption{Spectra retrieved by GMCA in our third and fourth toy models with a total number of counts corresponding to a $100$ ks observation. Left: Spectra retrieved by a GMCA with inpainting step of the two Gaussians in our toy model 3, with a first line emission-to-continuum ratio of 5.93 and a thinner binning ($14.6$ eV). Right: Spectra retrieved with an inpainting step of the two thermal emissions in our toy model 4, with a first thermal emission-to-continuum ratio of 2.64. To compare both error estimation methods, we added the error bars and the $kT$ given by 100 MC realizations, and those given by 100 block bootstraps resamplings of a single MC realization.}
\label{fig:modeles3_4}
\end{figure*}

    The error bars obtained through GMCA applied on 100 block bootstrap resamplings are slightly overestimated in comparison with those obtained with Monte-Carlo, but this does not have a crucial impact on the best-fit parameters obtained in \texttt{Xspec} (see Figure \ref{fig:modeles3_4}).

%-------------------------------------- Two column figure

\section{GMCA applied on toy models with more than two components}
\label{newtoys}

        We designed two more toy models featuring three sources instead of two (see Table \ref{table:Toy_model}, and Table \ref{table:ratios} for flux ratios). In our third toy model, we put a synchrotron and two Gaussians centered respectively on $6.54$ keV and $6.63$ keV. The one at $6.63$ keV has a norm equal to $0.7$ times that of the other one. Here, the Gaussians are the instrumental responses to a Dirac, hence they have a smaller width than in the first toy model. This is what we would get if the first wide Gaussian truly was the sum of two slightly shifted thinner ones. As we need a more precise definition in energy, the binning is thinner than in our previous toy models ($14.6$ eV), but the total number of counts is of the same order.
    
    In our fourth toy model, we input a synchrotron and two thermal emissions, one with $kT$ equal to $0.5$ keV, the other with $kT$ equal to $2$ keV. The norm of the first thermal emission is equal to $0.7$ times that of the second one. For the images, we used the blue- and redshifted Fe components shown in Fig. \ref{fig:real_fe}. As for our first two toy models, we added to our third and fourth toy models a flat image associated with the spectrum of an instrumental noise, and we generated Poisson noise on the whole data cube. The total number of counts of the synchrotron corresponds to a $100$ ks observation, and the second main component (brightest Gaussian or thermal emission)-to-synchrotron ratios we tested are the same as before.

        The GMCA is able to properly disentangle the three sources for the highest second-main-component-to-continuum ratios, but when the sources weaken, it only retrieves the synchrotron and a second source that is a composite of the two Gaussians, or of the two thermal emissions. Using the inpainting step helps to disentangle the three sources a little longer and improves the spectra in the thermal emission case, but the weakest thermal emission is underestimated: the leakage mechanism is more difficult to correct with three sources to disentangle than with only two of them. In Fig. \ref{fig:modeles3_4}, we can see an example of correct disentanglement of the components in both toy models. The presented line-to-continuum and thermal-to-continuum ratios are the last ones to give correct images and correct spectra for every component.
    
    We fitted the retrieved thermal emission spectra of our fourth toy model in \texttt{Xspec} in order to estimate $kT$. We first used as error bars the standard deviation of 100 MC realizations; we then took the standard deviation of 100 block bootstrap resamplings of a single MC realization. The  temperature of the first thermal emission is slightly overestimated with MC error bars, but the overestimation is of the same order as with our second toy model. However, this temperature is consistently retrieved with the block bootstrap error bars. The temperature of the second thermal emission is slightly underestimated in both cases.

\section{GMCA applied to real data}
\label{sect:real data}

\begin{figure}
{\includegraphics[width = 9cm]{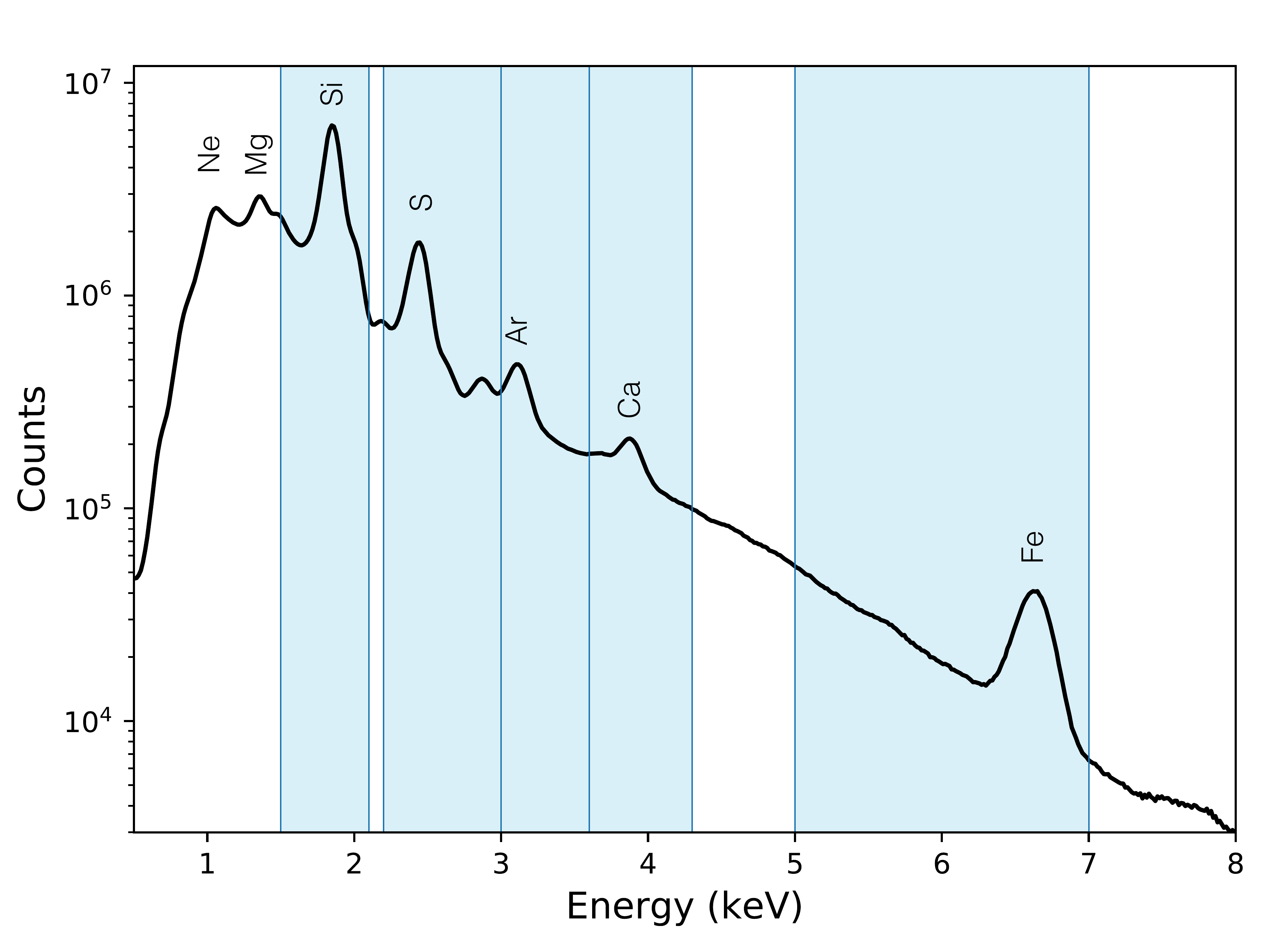}}\\
\vspace{-0.5cm}
\caption{Spectra of Cassiopeia A from the deep 2004 observations. The main emission lines are labeled as well as the energy ranges used for the GMCA algorithm in Fig.~\ref{fig:bluered}.  }
\label{fig:CasA_spec}
\end{figure}  

Following the consistency and the robustness
tests described above, we applied the GMCA to the deep {\it Chandra} observations of Cassiopeia A, which was observed with the ACIS-S instrument in 2004 for a total of 980 ks (ObsID : 4634, 4635, 4636, 4637, 4638, 4639, 5196, 5319, 5320). The spectrum from the whole SNR, together with the main emission features, is shown in Fig.~\ref{fig:CasA_spec}. The event lists from all observations were merged in a single data cube. For each application described in the sections below, the spatial and spectral binning were adapted so as to obtain a sufficient number of counts in each cube element. No background subtraction or vignetting correction has been applied to the data. We note that due to the lack of exposure and background map handling with the current version of GMCA, the method cannot yet be applied to a large mosaic of observations.

\subsection{Asymmetries of the Fe K distribution in Cassiopeia A}
\label{fek}

\begin{figure}
{\includegraphics[width = 9cm]{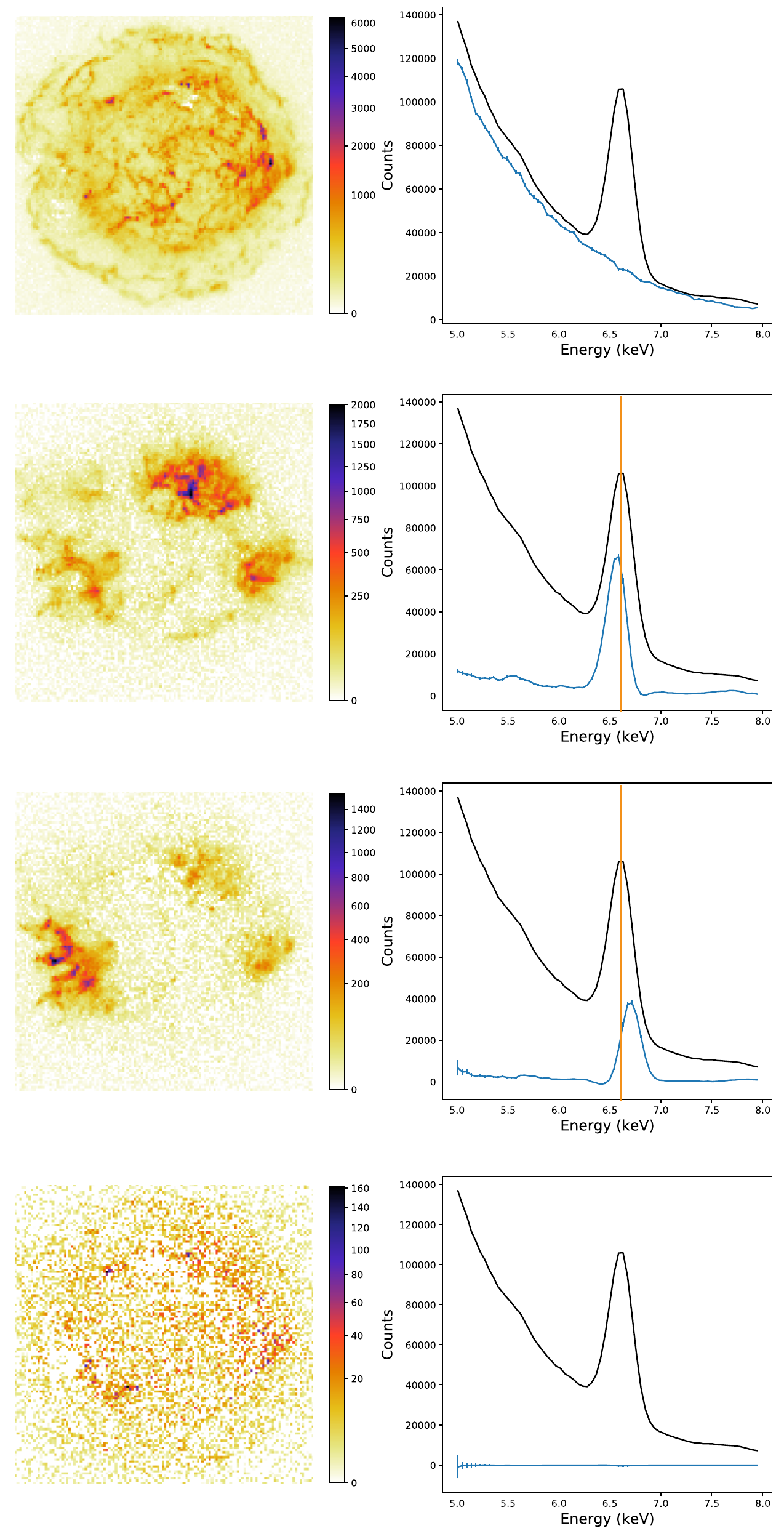}}\\
\vspace{-0.5cm}
\caption{Images and spectra retrieved by the GMCA with inpainting step in the real data from Cassiopeia A between $5$  and $8$ keV. The first source corresponds to a synchrotron emission, while the two following sources are parts of the Fe distribution; the first of these latter two is a redshifted part of this distribution, while the second is a blueshifted part of the distribution. The error bars in both parts of the Fe distribution are retrieved thanks to a block bootstrap with blocks of size 78. A fitting in \texttt{Xspec} gives the line energies of $6.726 \pm 0.002$ keV for the blueshifted part and $6.561 \pm 0.001$ keV for the redshifted part. The last image is the one we identify as noise.}
\label{fig:real_fe}
\end{figure}

        We first applied the GMCA to the Cassiopeia A observation between $5$  and $8$ keV, where the prominent features are known to be the synchrotron emission and the Fe K line complex. To allow for unexpected sources to be retrieved by the algorithm, it is recommended to decompose the data into a larger number of components than expected as a first guess. By doing this, we obtained three physically meaningful components in Cassiopeia A: continuum emission and two Gaussian lines that appear to be slightly shifted with respect to one another, and with respect to the Fe K average energy. The first component is undoubtedly the synchrotron emission, for which the image is coherent with our knowledge of its spatial distribution; the corresponding spectrum can be described as a power law (Fig. \ref{fig:real_fe}, top panel). The two other components have spectra corresponding to blue- and redshifted Fe line emission (Fig. \ref{fig:real_fe}, middle panels), and the associated images show clumps typical of the spatial distribution of Fe in Cassiopeia A \citep[see Fig.7 of ][]{delaney10}. If we instead require the algorithm to find only two components, it retrieves the synchrotron emission and a composite of the two Fe components. If require the algorithm to find more than three components, the additional retrieved sources are simply noise. The bottom panel of Fig. \ref{fig:real_fe} shows an image of what we identify as noise in such a case. 

        The block bootstrap resampling step outlined in Sect.~\ref{sec:boot} allowed us to extract the spectra corresponding to the different components above. Fitting the Fe K line emission in \texttt{Xspec} with a Gaussian model, the redshifted part was found to peak at $6.726 \pm 0.002$ keV, and the blueshifted part peaks at $6.561 \pm 0.001$ keV. These energies suggest a relative velocity between the red- and blueshifted components of $7440$ km~s$^{-1}$, a value that is coherent with the results shown in Fig.~7 of \cite{delaney10}.
    
%    and to fit t/fit them in XSPEC.
%    we retrieved the spectral error bars that allowed a fitting in Xspec of the Fe K lines. The red-shifted part was found to peak at
%$6.726 \pm 0.002$ keV and the blue-shifted one at $6.561 \pm 0.001$ keV . These energies suggest a relative velocity between the red and blue-shifted parts of $7440$ km~s$^{-1}$. This velocity is coherent with the results of \cite{delaney10}, Figure 7.

Our method allows direct imaging of the red- and blueshifted Fe K components with unprecedented spatial resolution. In addition, instead of estimating a mean shift in each line of sight (such as would be obtained when fitting with one Gaussian), our method can disentangle the red- and blueshifted components along a line of sight as shown in Fig. \ref{fig:real_fe}, where both emissions co-exist in the southeast.

\subsection{Spatial structures of the main line emissions in Cassiopeia A}

\begin{figure*}
\begin{center}
{\includegraphics[width = 15.8cm]{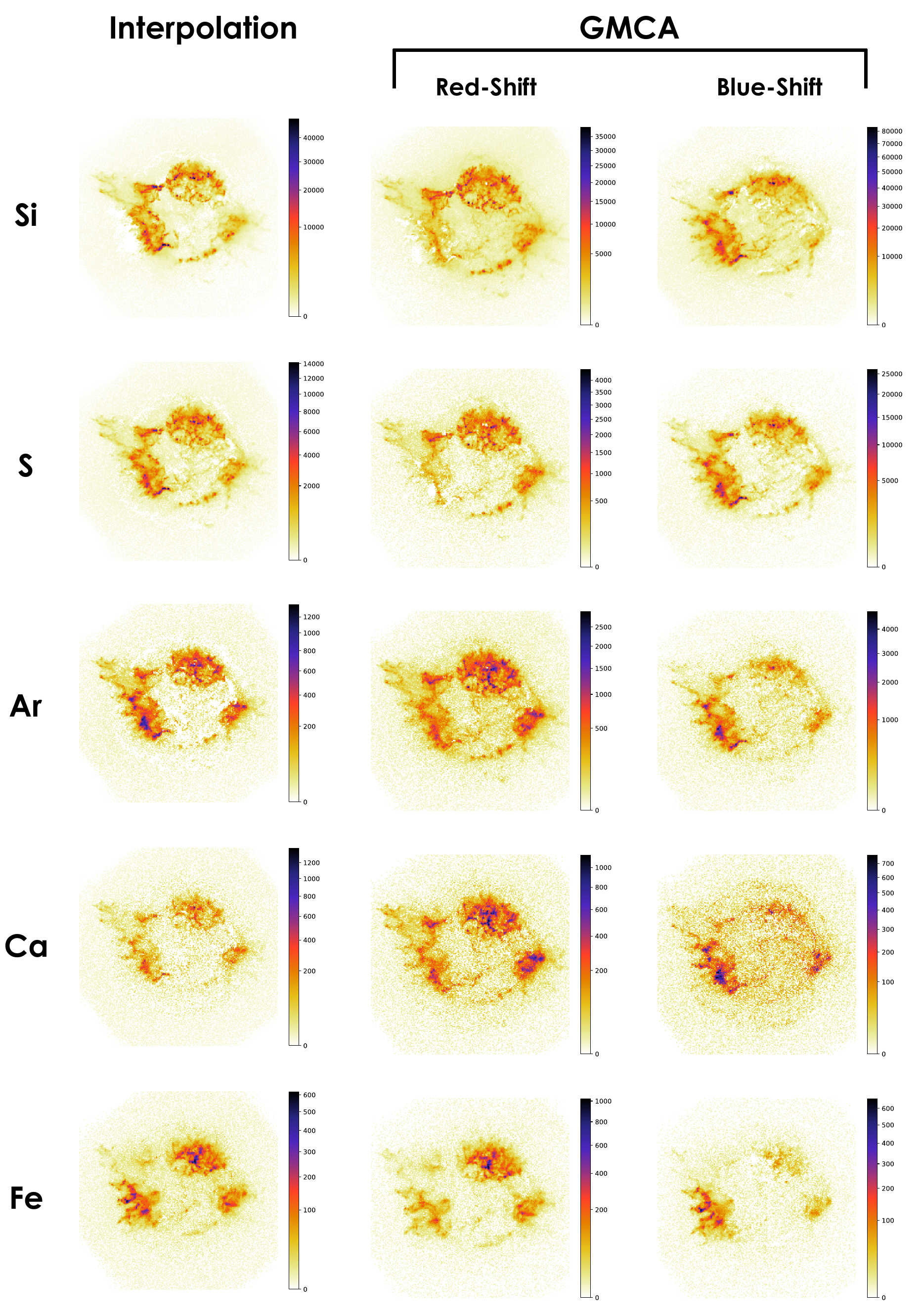}}\\
\vspace{-0.4cm}
\caption{Images of the spatial structure of the main line emission in Cassiopeia A as retrieved by an interpolation method (left column), and from application of GMCA around the respective line emission region (middle and right columns). In all cases, the GMCA algorithm decomposes the line emission into two images, corresponding to spectra that are slightly redshifted or slightly blueshifted with respect to the rest-frame line position. The energy ranges used for GMCA are shown in Fig.~\ref{fig:CasA_spec}. The energy ranges used for the interpolation method are respectively : $1.7 - 2$ keV, $2.25-2.6$ keV, $3-3.35$ keV, $3.7-4.1$ keV and $6.2-7.1$ keV.}
\label{fig:bluered}
\end{center}
\end{figure*}

        Figure~\ref{fig:bluered} shows an application of the method to the main line emission bands in Cassiopia A, centred on Si, S, Ar, Ca, and Fe. In each case, 
    %applied the same method around the main line emissions in Cassiopeia A, and in each case 
    the GMCA was able to retrieve two images corresponding to a slightly redshifted and a blueshifted component.%The components are individually barely visible in the overall spectra.} 
    
    We compared the spatial structures of these components to what could be retrieved by an interpolation method (see Section \ref{imagefidelity}) around these same line emissions. As we can see in Figure \ref{fig:bluered}, both methods give consistent results, although the GMCA retrieves more structures for faint lines (Ar and Ca). More importantly, the GMCA can probe structures within a broad line and reveal line shifts, information that cannot be yielded by the interpolation method. The blueshifted and redshifted images in Si, S, Ar, and Ca are very similar (but differ from Fe). This attests to the robustness of GMCA, because the energy ranges are completely independent.

\subsection{Spatial distribution of continuum components in Cassiopeia A}

\begin{figure*}
\begin{center}
{\includegraphics[width = 18.3cm]{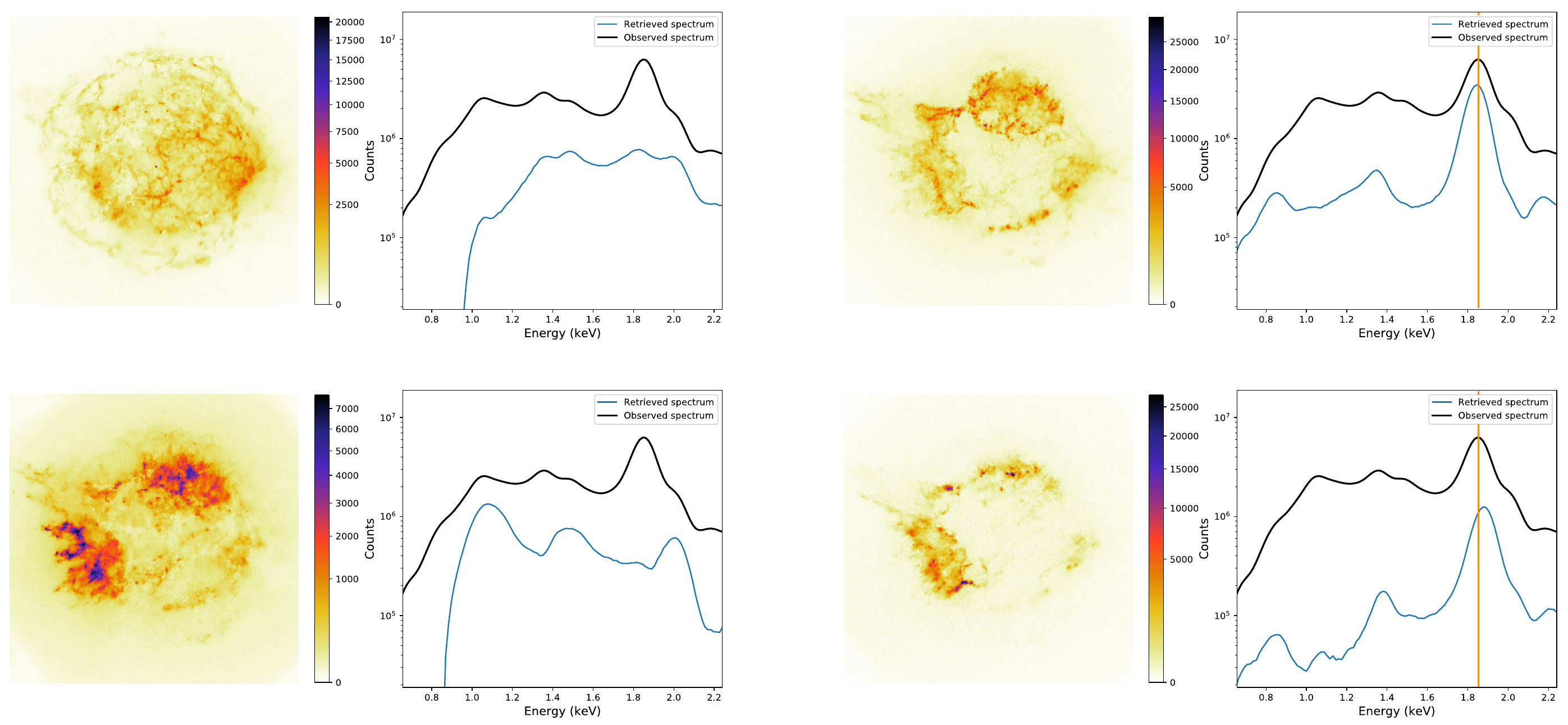}}\\
\vspace{-0.3cm}
\caption{Images and spectra retrieved by the GMCA with inpainting step in the real data from Cassiopeia A between $0.6$ keV and $2.3$ keV with a spectral binning of $14.6$ eV and pixels of a $0.9$ arcsec size. On the upper left, we recognize the synchrotron emission. On the lower left, the retrieved component seems to be dominated by Fe L. On the right, we see two components dominated  by red- and blueshifted Si, respectively.}
\label{fig:real_si}
\end{center}
\end{figure*}

        We applied our method on Cassiopeia A data between $0.6$ and $2.3$ keV. The number of counts being higher in this energy band, we used data with a finer spectral binning ($14.6$ eV instead of $43.8$ eV) and smaller pixels ($0.9$ arcsec instead of $1.8$ arcsec).
    
    Figure~\ref{fig:real_si} shows that the four components were retrieved. The first corresponds to the synchrotron emission, and is coherent with the image we retrieved between $5$ and $8$ keV. It is the first time an image of the synchrotron has been extracted in these energy bands, where it is dominated by the ejecta emission. Such a map of the low-energy synchrotron emission is very valuable for the study of the energy dependence of the synchrotron rim width. A second component has a spatial distribution highly similar to that of the Fe K between $5$ and $8$ keV, and the line emission complex at $\sim$ 1 keV in the corresponding spectrum seems to indicate that this component may be dominated by the Fe L complex.
    
    The final two components have spectra corresponding to slightly red- or blueshifted Si emission. Their spatial distributions are similar: we can thus deduce that both components correspond to Si emission, one being slightly redshifted, and the other slightly blueshifted. The morphology of the two parts is globally consistent with previous works but is endowed with more details (see e.g., Figure~7 of \cite{willingale02}, or \cite{delaney10} for a comparison with optical images). We note that each thermal component is not completely dominated by a unique line structure. For example, we see that the Si components (Figure \ref{fig:real_si}, right panels) also contain oxygen and magnesium emission in their spectra. Oxygen and magnesium are grouped together with Si by the algorithm because they have similar plasma conditions (temperature, abundances, ionization stage) and spatial distributions.
More surprisingly, the component exhibiting Fe L emission (Figure \ref{fig:real_si} bottom left panel) also has strong Mg XII and Si XIV line emission. This indicates that the Fe L is co-spatial with Mg and Si in a higher ionization state than in the Si-dominated components. While the reason for this difference is still unclear, this example shows the power of GMCA to disentangle  physical components in complex environments.

\section{Discussion and Conclusions}
\label{conclusion}

        The separation of entangled components in the X-ray data of extended sources is a challenging task. Isolation of the morphology and associated spectrum of the individual components could provide new insight into the physical and thermodynamical conditions of the plasma in these objects.
%    that could provide plenty of physical information by disentangling detailed individual spatial maps and spectra of the different X-ray emission sources. 
    In the case of supernova remnants, those measurements could lead to a better understanding of the explosion mechanisms, gas heating, and particle acceleration. 
    
     Here we present a method based on the GMCA, a blind source-separation algorithm developed to extract the CMB from {\it Planck} data. The method uses all of the information contained in data cubes $(E,x,y)$, and extracts the unique spatial and spectral signatures of the entangled components without any prior information (neither physical models nor instrument response functions). It has been applied here to X-ray data for the first time, and we have shown that it provides better results than the usual methods in use in this field. 
    
    The GMCA needs to be applied to data with a large total number of counts. When such data are available, it can successfully disentangle highly spatially correlated sources, as was shown with our toy models (Sections \ref{toy} and \ref{newtoys}). A first application to real {\it Chandra} data of Cassiopeia A in different energy bands, detailed in Section \ref{sect:real data}, gave promising results, highlighting the asymmetries in the Si, S, Ar, Ca, and Fe K spatial distributions by retrieving two maps associated to spectra that are slightly red- or blueshifted with respect to the rest-frame line.
   
        The main conclusions of our study are the following:
   
\begin{itemize}
\item \textbf{Morphological fidelity:} In every example we tested, it appeared that the GMCA yields accurate images of the sources it retrieves, very close to the original ones we injected in the toy model. Furthermore, while the cases we tested were very challenging, the sources being spatially highly entangled, our method succeeded in retrieving detailed disentangled images of each component. Lastly, the algorithm never retrieved any artifact that did not belong to the toy model: when the second-component-to-main-component ratio was too weak, the second component was not retrieved, but everything that was retrieved could be trusted was a bona-fide component, and not a false detection.

\item \textbf{Spectral fidelity:} While the initial GMCA retrieves correct images, there is a leakage which affects parameters that depend on a wide energy range when the spectra are fitted in \texttt{Xspec}. An inpainting step that we added after the internal loops of the GMCA corrected most of the overestimation of the spectrum caused by the leakage, and improved the retrieved temperatures.

\item \textbf{Block bootstrap:} Spectral analyzing tools such as \texttt{Xspec} need error bars in order to fit physical models. The block bootstrap resampling method tested here is a promising way to estimate error bars from a single set of data.

\item \textbf{Performance:} The ability of the GMCA to disentangle components depends on the total number of counts in the data, on the number of counts of each component, and on the nature of the data itself: performance is very case-specific. In this paper, we focused on the study of highly spatially entangled sources, which are frequent in the study of SNRs, and represent an extremely challenging analysis task. For that reason, the weakest ratio at which every component can be successfully retrieved depends on the morphological and spectral diversity. We also note that  the algorithm is more successful in finding faint features when applied to narrow, targeted energy bands rather than when applied to the full energy range. To conclude, GMCA is a fast-running algorithm, taking only a few minutes to extract sources from a $200*200*300$ data cube on a single-core personal computer.

\end{itemize}

The version of GMCA we used in this study was originally developed to handle the Gaussian noise in {\it Planck} data. The method will be enhanced in future work by inclusion of a treatment for Poisson statistics that should help to retrieve fainter components and diminishing leakages. In addition, exposure and background cubes will be implemented for application of the method to large mosaic observations. The use of physically motivated spectral models to guide the component-separation process could also be envisaged.
%Providing spectral constraints using \gwp{physically-motivated models} could also be used to improve spectral accuracy.

New spectro-imaging instruments with increased effective area and high spectral resolution, such as Athena, will provide data whose tremendous potential cannot be fully exploited with  existing data analysis methods. The GMCA provides a new way to leverage all possible dimensions in the data, thus allowing a maximum of physical information to be obtained.

\begin{acknowledgements}
This research made use of Astropy,\footnote{http://www.astropy.org} a community-developed core Python package for Astronomy \citep{astropy:2013, astropy:2018}. GWP acknowledges funding from the European Research Council under the European Union's Seventh Framework Programme (FP7/2007-2013)/ERC grant agreement No. 340519. JB acknowledges funding from the European Community through the grant LENA (ERC StG - contract no. 678282).

\end{acknowledgements}

   \bibliographystyle{aa} % style aa.bst
   \bibliography{article-GMCA} % your references Yourfile.bib

%\begin{thebibliography}{}
%
%   \bibitem[Zheng(1997)]{zheng} Zheng, W., Davidsen, A. F., Tytler, D. \& Kriss, G. A.
%      1997, preprint
%\end{thebibliography}

\begin{appendix} %First appendix

\section{Wavelets and starlets}
\label{sect:wavelets}
        
    A wavelet is a square-integrable function of zero mean. Briefly, a wavelet transform consists in contracting a mother wavelet and convolving it with an image, each scale providing a new image. Each wavelet scale contains information about structures of a specific size. For that reason, a wavelet transform proves useful to disentangle components using their morphological specificities. The starlet transform is a special case of bi-dimensional wavelets, which have been specifically designed to efficiently represent isotropic structures in images. Therefore, this particular case of wavelets has proven to be well-adapted to analyzing astrophysical images.
    
    %The Starlet transform decomposes a $n*n$ image $c_0$ into a coefficient set $W=\{ w_1, ... , w_J, c_J\}$, as a superposition of the form:
    
    %\begin{equation}
    %c_0[k,l]=c_J[k,l]+\overset{J}{\underset{j=1}{\sum}}w_j[k,l]
    %\end{equation}
    
    %where $c_J$ is a coarse or smooth version of the original image $c_0$ and $w_j$ represents the details of $c_0$ at scale $2^{−j}$. Thus, the algorithm outputs $J+1$ sub-band arrays of size $n*n$. (The present indexing is such that $j=1$ corresponds to the finest scale or high frequencies).
    
    The starlet transform first builds a sequence of approximations of an $n*n$ image $c_0$ at increasingly large scales $\{ c_1, \cdots,c_J\}$. Each approximation is obtained from the previous one through a convolution with a mother wavelet filter $\bar{h}^{(j)}$ at scale $j+1$:
    \begin{equation}
    c_{j+1}[k,l]=(h^{(j)} \star c_j)[k,l],
    \end{equation}
    where the filter $h^{(0)}$ is defined as:
    
%   \begin{equation}
%   h_{2D}=\begin{pmatrix} 
%1/16 & 1/4 & 3/8 & 1/4 & 1/16 
%\end{pmatrix}
%\begin{pmatrix} 
%\frac{1}{16} \\ 
%\frac{1}{4} \\ 
%\frac{3}{8} \\ 
%\frac{1}{4} \\ 
%frac{1}{16} 
%\end{pmatrix}
    %\end{equation}
    
\begin{equation}
  h^{(0)}=\frac{1}{256}\begin{pmatrix} 
1 & 4 & 6 & 4 & 1 \\
4 & 16 & 24 & 16 & 4 \\
6 & 24 & 36 & 24 & 6 \\
4 & 16 & 24 & 16 & 4 \\
1 & 4 & 6 & 4 & 1 
\end{pmatrix}  
.\end{equation}
According to the "\`a trous" algorithm, consecutive filters,  $h^{(j)}$ , are obtained by adding zeroes between the nonzero filter elements so as to dilate the filter by a factor of two from scale to scale \cite{starck_murtagh_fadili_2015}.

    The wavelet coefficient at scale $j+1$ is then defined as the difference between consecutive large-scale approximations:
    \begin{equation}
    w_{j+1}[k,l] = c_j[k,l] - c_{j+1}[k,l]
    .\end{equation}
    This eventually yields a decompostion of the image $c_0$ into a coefficient set $W=\{ w_1, ... , w_J, c_J\}$.
    
The reconstruction of the initial image $c_0$ is then obtained by a simple coaddition of all wavelet scales and the final smooth sub-band:

    \begin{equation}
    c_0[k,l]=c_J[k,l]+\overset{J}{\underset{j=1}{\sum}}w_j[k,l]
    .\end{equation}

\begin{figure*}
\centering
\includegraphics[width = 19cm]{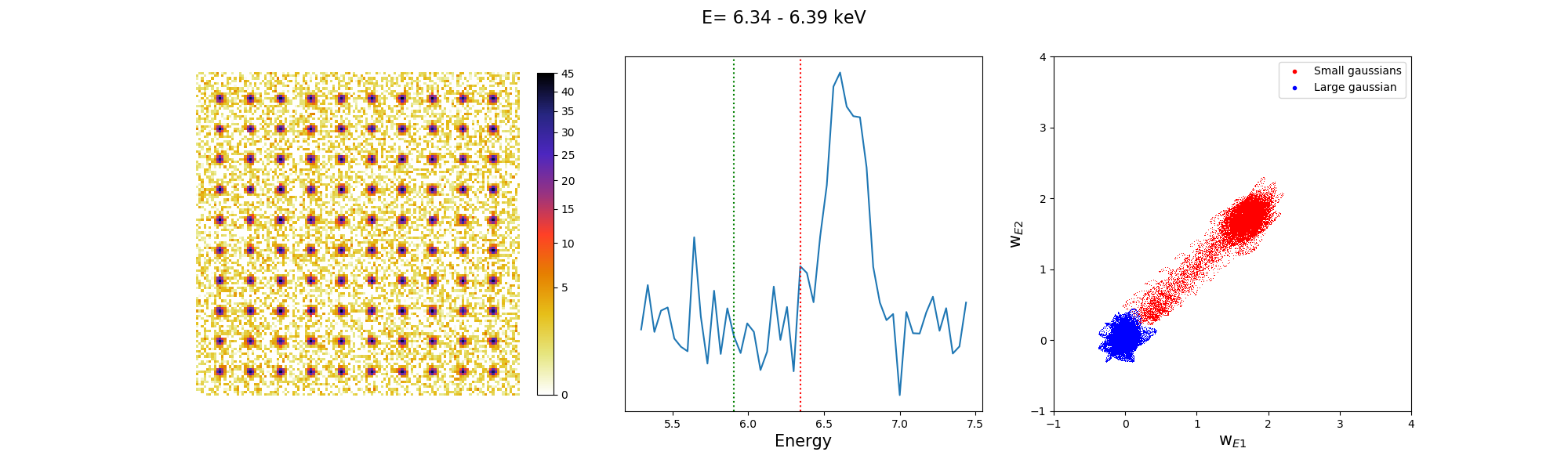}
\includegraphics[width = 19cm]{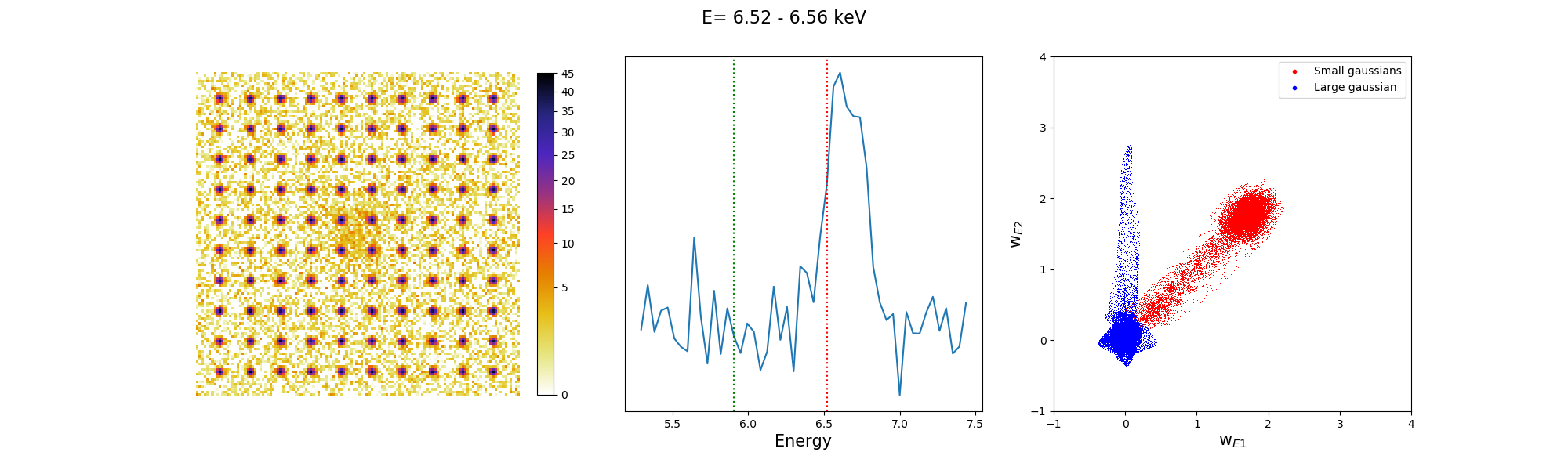}
\includegraphics[width = 19cm]{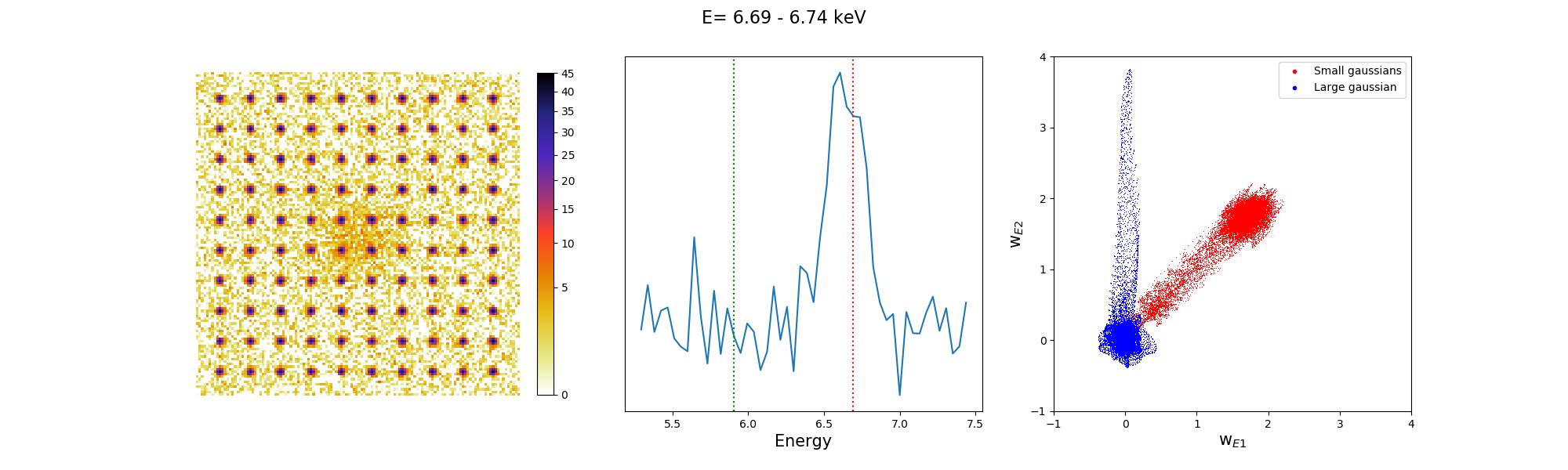}
\caption{Illustration of the relevance of wavelet transforms to separate components in a cube $(E,x,y)$ using a simple toy model. Left: Slice of the cube at the current energy $E2$. Center: total spectrum. The green line corresponds to $E1$ while the red one corresponds to $E2$. Right: Representation of the coefficients of both components in the third wavelet scale at $E2$ as a function of the coefficients of both components in the third wavelet scale at $E1$. The coefficients of the two components are clearly dissociated in the wavelet space, whereas their images are tightly entangled. }
\label{fig:wavelet_}
\end{figure*}

    In Figure \ref{fig:wavelet_}, we build a very simple toy model to show the relevance of starlet transforms to separate components in a cube $(E,x,y)$. The data cube is the sum of two components: an array of small spatial Gaussians multiplied by a flat spectrum and a large spatial Gaussian multiplied by a spectral line. A Gaussian noise with a standard deviation of $2$/pixel is added to the cube. The figure points out the differences between the coefficients of the two components in the third starlet scale.

\section{Evaluating the number of components to retrieve. }
\label{sect:numbcomp}

From a statistical viewpoint, evaluating the number of components to be retrieved boils down to a model-selection problem. Testing for an increased number of components $n$ is equivalent to testing a model with $n_{x}*n_{y}+n_{E}$ additional degrees of freedom and selecting the one with the lowest Akaike information criterion (AIC).

\begin{figure*}
\centering
\includegraphics[width = 19cm]{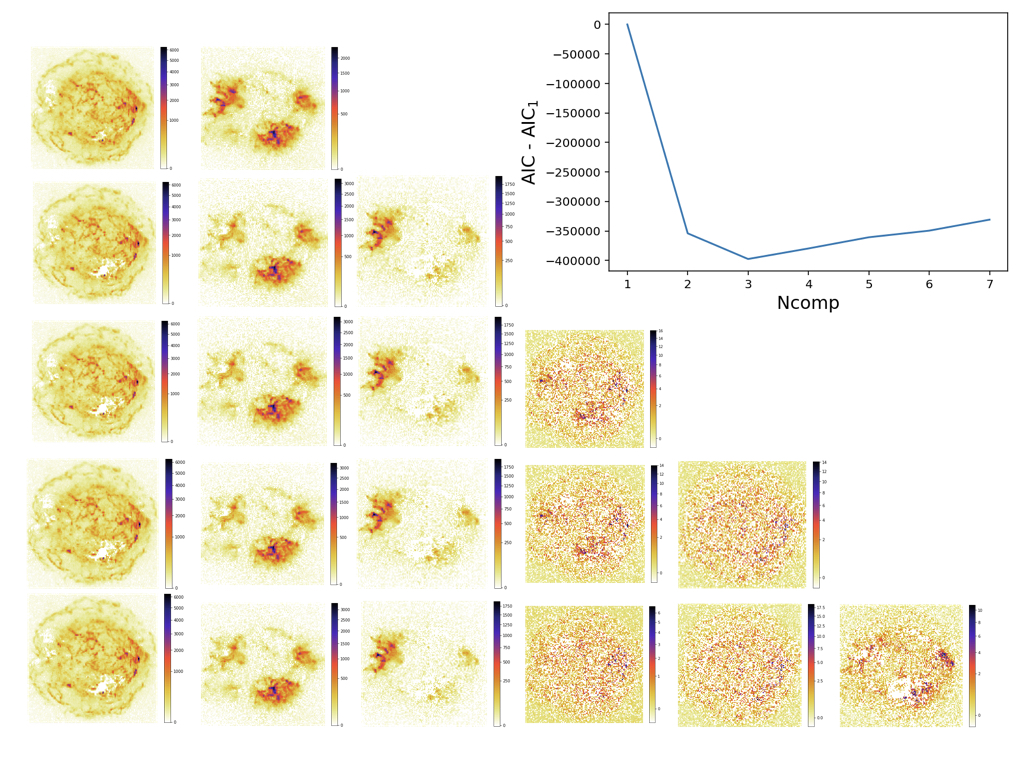}
\caption{Components retrieved by the GMCA applied on Cassiopeia A real data for different values of $n$. On each line, the retrieved components corresponding to a certain $n$, beginning with $n=2$ on top. On the right, the AIC of the model as a function of this number of components $n$.
}
\label{fig:numb_examples}
\end{figure*}

For a given model, the AIC is defined as twice the difference between the log-likelihood and a complexity term $p$ :

\begin{equation}
AIC=-2\log(L)+2p
,\end{equation}

where $p=(n_{x}*n_{y}+n_{E})*n$ is the number of degrees of freedom, and $L$, the log-likelihood for $n$ components, is defined as

\begin{equation}
L =  AS - AS\log(X)
,\end{equation}

with $A$ and $S$ being the solutions obtained by GMCA for $n$ components. 
For example, in the test presented in Figure \ref{fig:numb_examples} we applied the GMCA to Cassiopeia A real data between $5$ and $8$ keV with a number of components increasing from $1$ to $7$ and looked at the AIC of the different models. The images shown seem to indicate that after three components, no other meaningful components are retrieved, which is confirmed by the AIC.

It is important to note that if the couple $(A, S)$ were the maximum likelihood estimate, taking the AIC minimum would be a reliable criterion to determine $n$.
However, in the case of the GMCA algorithm, $(A, S)$ is not exactly a maximum likelihood since the GMCA algorithm makes use of a sparse regularization, which eventually yields solutions that do not maximise the likelihood. Furthermore, component separation problems are nonconvex and algorithms such as GMCA are only guaranteed to converge to a local minimum, which does not necessarily correspond to a global minimizer. 

Even if the test was satisfying in the example presented above, we cannot insure that it will be with any other data set for the reason we mentioned. In practice, the main components are stable and well retrieved for different values of n, but there can be fluctuations in the remaining noisy images that would impact the statistical tests even if no actual physics could be gleaned from their interpretation. Also, adding a physically meaningful component presenting a coherent structure, but too faint to have a clear statistical impact, may not minimize the AIC. Hence, the AIC can be a useful figure of merit to confirm the relevance of a certain chosen $n$, but should not replace a human interpretation nor be directly implemented inside of the algorithm as an automatized selector for the number of components.

\section{Spatial and spectral accuracy}

\begin{figure*}
\centering
\includegraphics[width = 18cm]{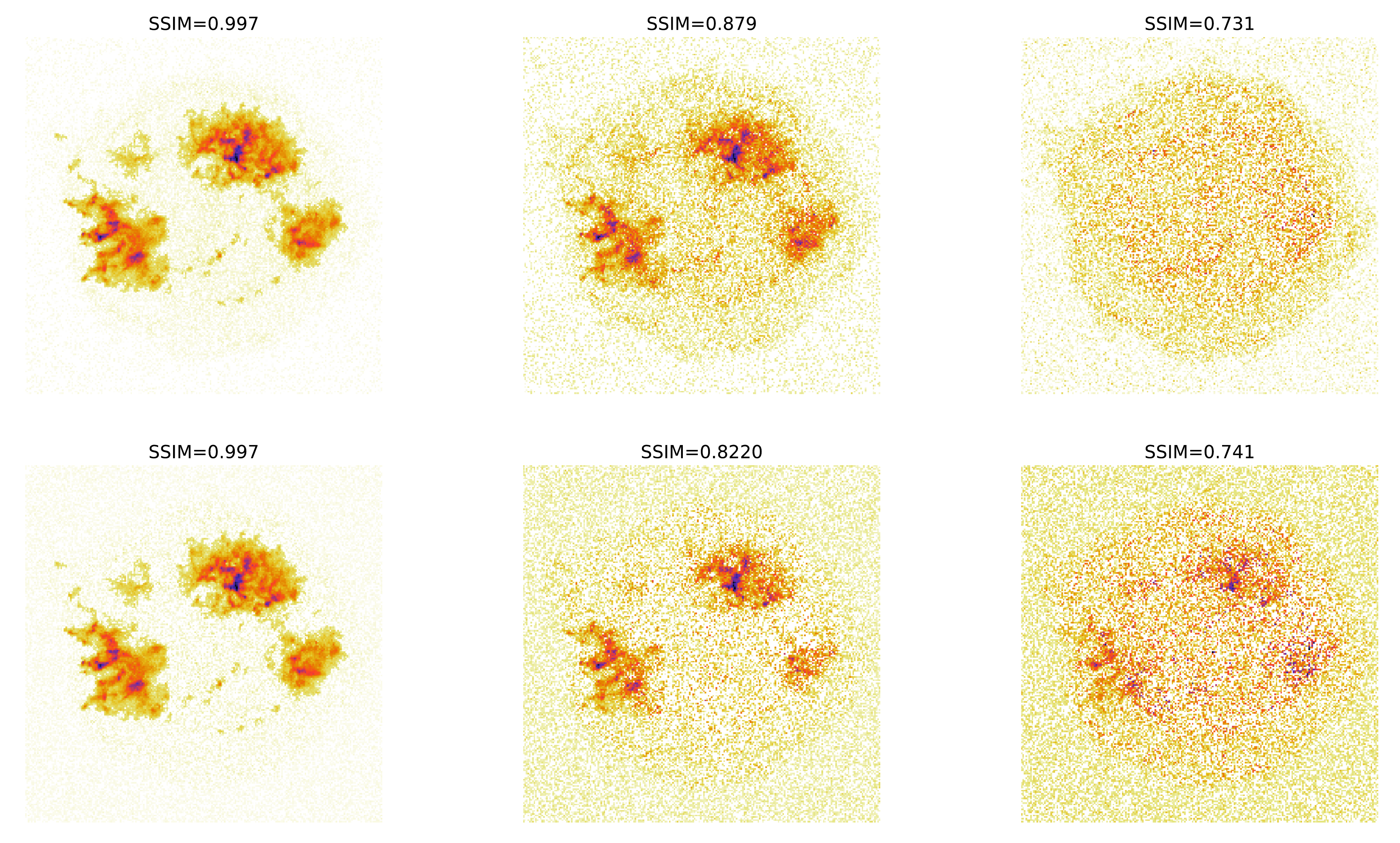}
\caption{Images of the Fe spatial structures in our first toy model as found by GMCA without fixing the Fe spectral shape (on top) and by an interpolation method (below) for the three Fe line-to-synchrotron ratios marked by arrows in Figure \ref{fig:ssim_ratios}). The SSIM coefficients are written on top of the images. Coefficients under $0.75$ describe images where the Fe structures are not recognizable, but the SSIM is still high because of the similarities between intrinsic Fe and synchrotron distributions.}
\label{fig:ssim_examples}
\end{figure*}

\begin{figure*}
  \subfloat{\includegraphics[width = 9cm]{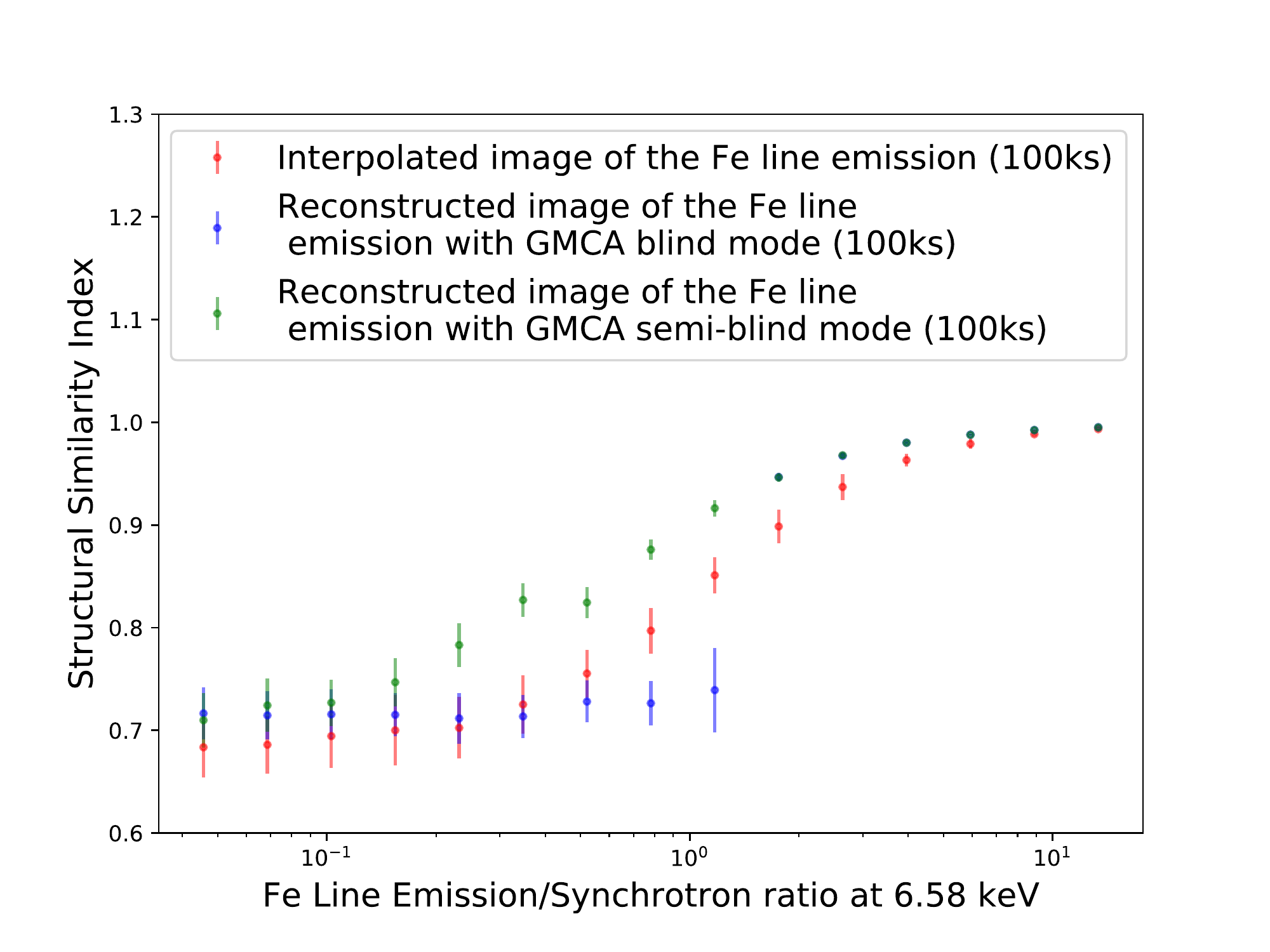}}
\subfloat{\includegraphics[width = 9cm]{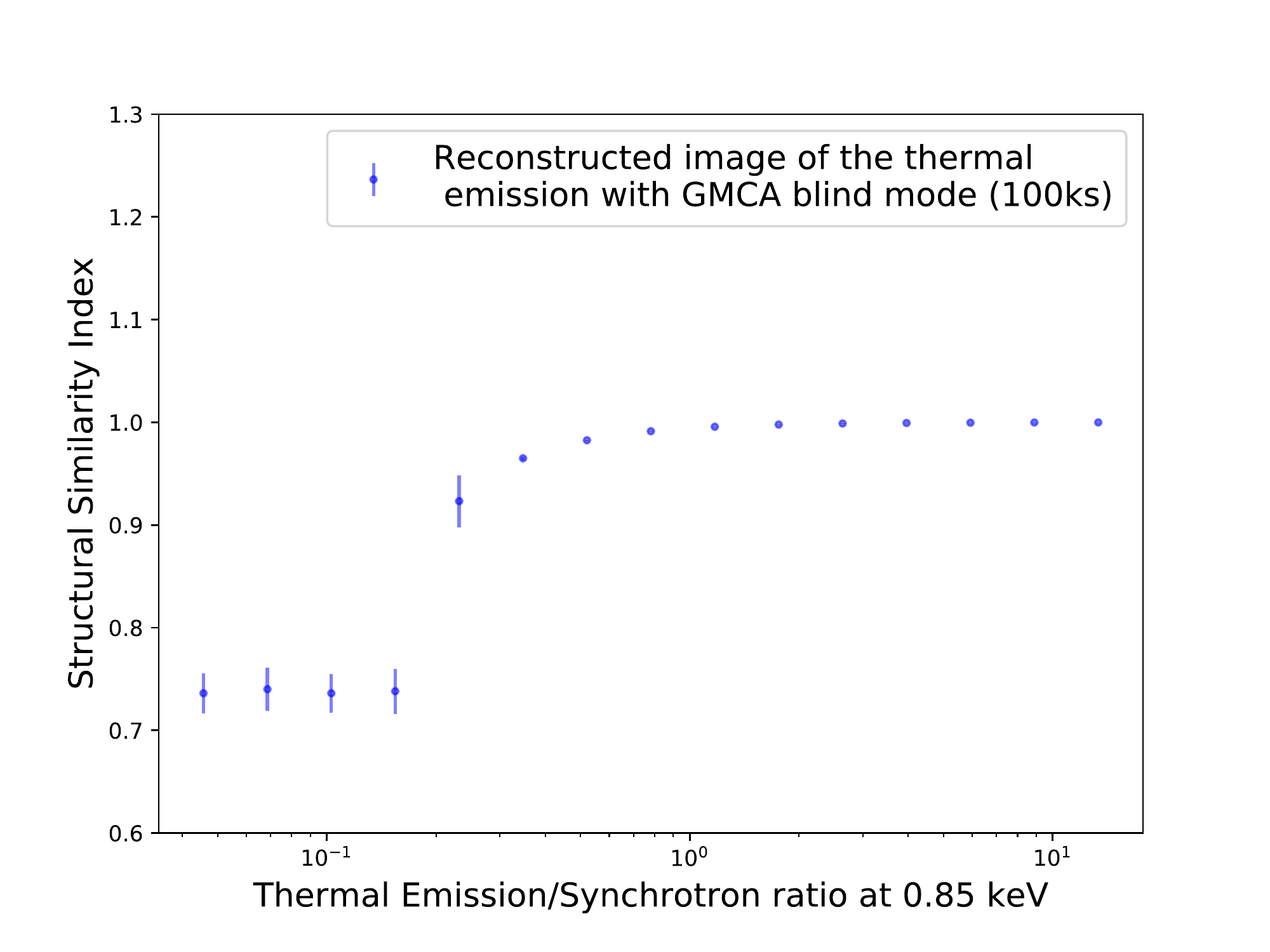}}
\caption{SSIM coefficients of the input and output images found by GMCA for a total number of counts corresponding to a $100$ ks observation. The points are the average of all Monte-Carlo realizations at a particular ratio, and the error bars the standard deviation of those realizations. Left: Comparison of the image quality obtained in retrieving the Fe structure in our first toy model for different line emission-to-synchrotron ratios, between an interpolation method, a GMCA in blind mode, and a GMCA in semi-blind mode. Right: Image quality of the thermal emission structure retrieved for different ratios by a GMCA in blind mode.}
\label{fig:ssim_ratios2}
\end{figure*}

\begin{figure*}
  \subfloat{\includegraphics[width = 9cm]  {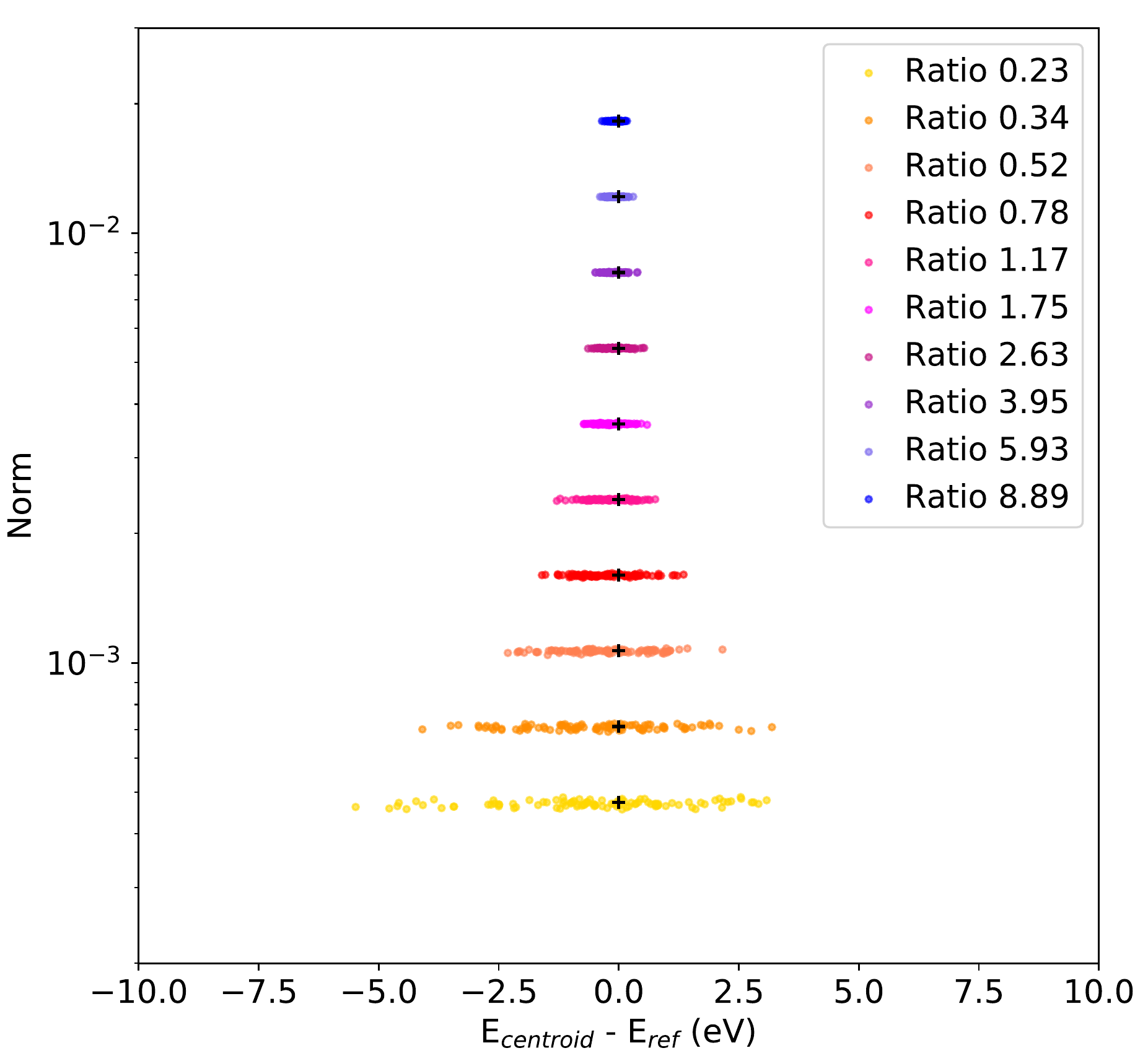}}
\subfloat{\includegraphics[width = 9cm]{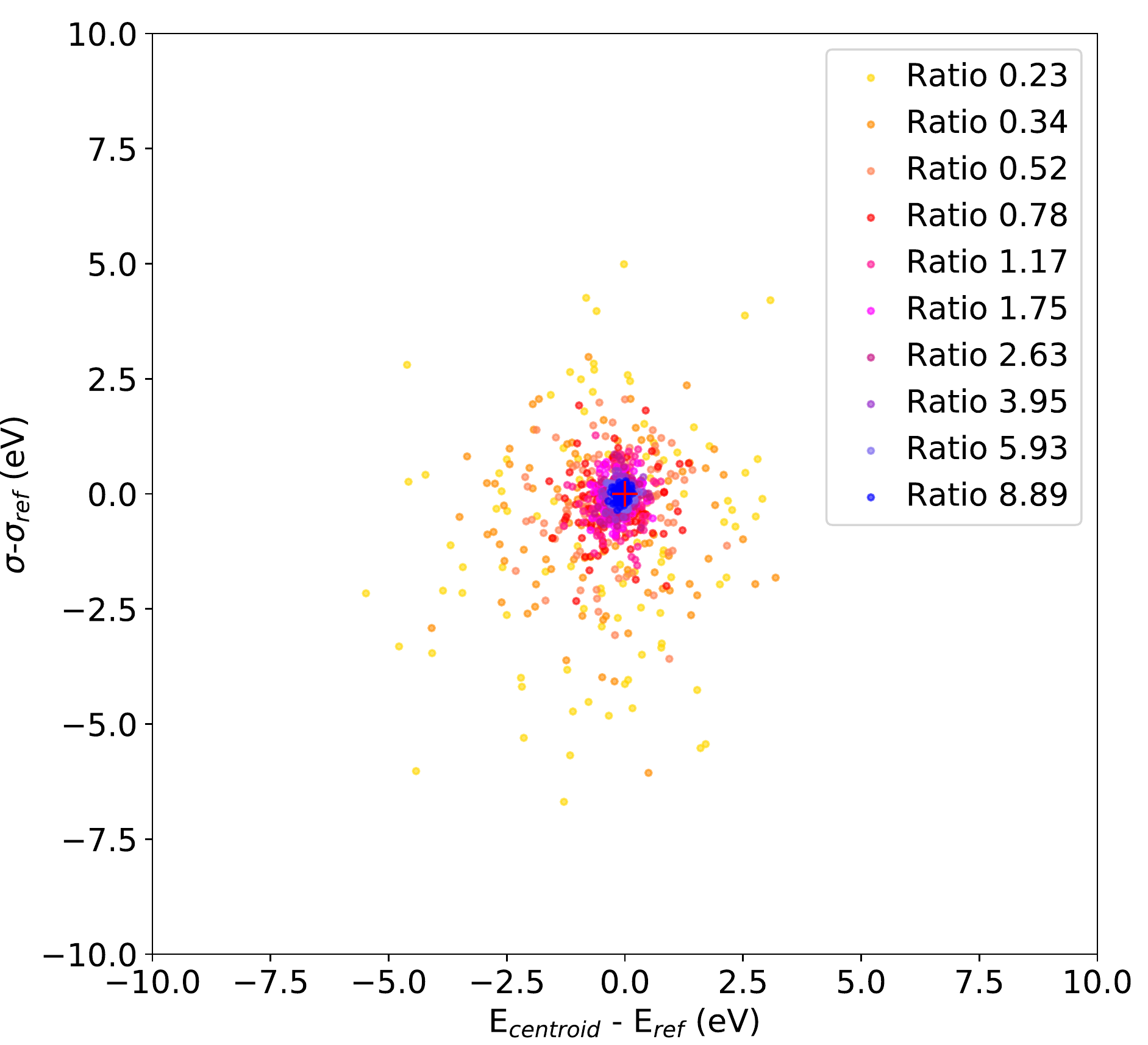}}
\caption{Parameters of the Fe K Gaussian as retrieved by \texttt{Xspec} on the  total spectrum of our first toy model for 100 realizations of each out of ten different Fe line-to-synchrotron ratios with a total number of counts corresponding to a $1$ Ms observation. Left: Retrieved $E_c$ and $\sigma$. Right: Retrieved norm and $E_c$. In both cases, the theoretical results are represented by black crosses.}
\label{fig:spectra_noGMCA}
\end{figure*}

In this section we present some additional figures resulting from our tests of the GMCA on our first two toy models. Figure \ref{fig:ssim_examples} shows examples of images of the Fe spatial distribution in our first toy model by GMCA and an interpolation method for three different Fe line-to-synchrotron ratios, and the corresponding SSIM coefficients. Figure \ref{fig:ssim_ratios2} shows the evolution of the accuracy of the retrieved images for $15$ ratios in our first and second toy models with a total count corresponding to a $100$ ks observation. Figure
\ref{fig:spectra_noGMCA} shows the parameters of the Fe K Gaussian in our first toy model as retrieved by \texttt{Xspec} without using GMCA. This latter offers a good comparison with Figure \ref{fig:model1}, where the parameters were retrieved by fitting the spectra given by GMCA.

\end{appendix} %First appendix

\end{document}